\documentclass[letterpaper,twocolumn,table,10pt]{article}
\usepackage[table]{xcolor}
\usepackage{usenix}

\usepackage{amsfonts,amsmath,amssymb,amsthm}
\usepackage{cleveref}
\usepackage{hyperref}
\usepackage{xurl}
\usepackage{subfiles}
\usepackage[switch]{lineno}
\usepackage{listings}
\usepackage{tabularx}
\usepackage{booktabs}
\usepackage{subcaption}
\usepackage{caption}
\usepackage{xspace}
\usepackage{soul}
\usepackage{mdframed}
\usepackage{enumitem}
\usepackage[linesnumbered,ruled,noend]{algorithm2e}
\usepackage{algpseudocode}
\usepackage{multirow}
\usepackage[compress]{cite}
\usepackage{makecell}
\usepackage{calc}
\usepackage{float}
\usepackage{tikz}
\usetikzlibrary{shapes,arrows,positioning,calc,decorations,matrix,3d,patterns,arrows.meta,decorations.markings,fit,shapes.callouts,backgrounds,shapes.misc,shadows.blur,decorations.pathreplacing}
\usetikzlibrary{fadings} 

\usepackage[available,functional,reproduced]{usenixbadges}

\usepackage{etoolbox}
\makeatletter
\patchcmd{\@algocf@start}%
  {-1.5em}%
  {0pt}%
  {}{}%
\makeatother

\tikzset{algpxIndentLine/.style={draw=black!15,very thin}}

\newcommand{\datastore}{\mathtt{D}}

\newcommand{\tble}{\mathtt{T}}

\newcommand{\pred}{\mathtt{P}}
\newcommand{\attr}{\mathtt{atr}}
\newcommand{\qry}{\mathtt{Q}}
\newcommand{\TimeOracle}{\textsc{TimeTest}}
\newcommand{\ExtendTuples}{\textsc{ExtendTuples}}
\newcommand{\BinSrchNextVals}{\textsc{BinSrchNextVals}}
\newcommand{\vpart}{v_\mathrm{partial}}

\newcommand{\Cand}{\textsf{test}}
\newcommand{\Canddoc}{\doc_\Cand}
\newcommand{\Cands}{s_\Cand}
\newcommand{\Ctrl}{\textsf{ctrl}}
\newcommand{\Ctrldoc}{\doc_\Ctrl}
\newcommand{\Ctrls}{s_\Ctrl}
\newcommand{\Ctrlr}{ r_\Ctrl}
\newcommand{\Ctxr}{r_\textsf{ctx}}
\newcommand{\Concat}{\mathbin{\Vert}}

\newcommand{\idfprefix}{\idf_\mathrm{prefix}}
\newcommand{\dfprefix}{\df_\mathrm{prefix}}
\newcommand{\doc}{\mathtt{d}}
\newcommand{\corp}{\mathcal{D}}
\newcommand{\Analyze}{\mathtt{analyze}}

\newcommand{\ngramshat}{\hat{W_n}}
\newcommand{\What}{\hat{W}}
\newcommand{\chars}{\textsc{chars}}
\newcommand{\score}{\mathrm{score}}
\newcommand{\tf}{\mathrm{tf}}
\newcommand{\df}{\mathrm{df}}
\newcommand{\pf}{\mathrm{pf}}
\newcommand{\phrase}{\mathrm{phr}}
\newcommand{\expand}{\mathrm{exp}}
\newcommand{\idf}{\mathrm{idf}}
\newcommand{\idfmpp}{\mathrm{idf}_\mathrm{mpp}}
\newcommand{\dfmpp}{\mathrm{df}_\mathrm{mpp}}

\newcommand{\scorebm}{\mathrm{score}}

\newcommand{\ngram}{\texttt{\_ngram}}
\newcommand{\spc}{\text{\textvisiblespace}}

\newcommand{\qtime}{\mathsf{time}}

\newcommand{\rndterm}{\textsc{freshterm}}
\newcommand{\Write}{\textsc{write}}

\newcommand{\boost}{\mathsf{boost}}
\newcommand{\dl}{\mathsf{dl}}
\newcommand{\avgdl}{\mathsf{avgdl}}
\newcommand{\match}{\texttt{match}\xspace}
\newcommand{\matchphraseprefix}{\texttt{match\_phrase\_prefix}\xspace}
\newcommand{\mpp}{\texttt{mpp}\xspace}

\newcommand{\push}{\textsc{push}}
\newcommand{\pop}{\textsc{pop}}

\newcommand{\heading}[1]{
\vspace{3pt}\noindent\textbf{{#1}}}

\newlength{\saveparindent}
\newlength{\saveparskip}
\usepackage{tcolorbox}
\tcbset{
  colback=black!10,
  colframe=white!50!black,
  boxrule=0.25pt,
  boxsep=4pt,
  left=3pt,
  right=3pt,
  top=3pt,
  bottom=3pt,
  arc=1mm,
}

\let\oldtwocolumn\twocolumn
\renewcommand\twocolumn[1][]{%
   \oldtwocolumn[{#1}{
   \begin{center}
          \vspace{-23pt}
          \newcommand*\circled[1]{\tikz[baseline=(char.base)]{
            \node[shape=circle,draw,inner sep=1pt] (char) {#1};}}

\definecolor{figVisible}{HTML}{4E7D17} %
\definecolor{figHidden}{HTML}{A32D2D}  %
\definecolor{figRLS}{HTML}{0F6E56}     %
\definecolor{figDLS}{HTML}{9A5B0C}     %
\definecolor{figChip}{HTML}{2C2C2A}    %
\centering

\begin{tikzpicture}[
    font=\footnotesize,
    >=Latex,
    x=1cm, y=1cm,
    panel/.style={
        draw=none,
        rounded corners=6pt,
        inner sep=6pt,
        minimum height=2.35cm,
        align=left
    },
    badge/.style={
        circle,
        draw=black!35,
        fill=white,
        line width=0.35pt,
        minimum size=4.2mm,
        inner sep=0pt,
        font=\footnotesize\bfseries
    },
    lbl/.style={font=\footnotesize},
    arrow/.style={->, line width=0.8pt, draw=black!65},
    thin/.style={->, line width=0.6pt, draw=black!55},
    pill/.style={
        draw=black!25,
        fill=white,
        inner sep=2.4pt,
        font=\footnotesize
    },
    box/.style={
        draw=black!45,
        fill=white,
        rounded corners=3pt,
        line width=0.5pt,
        inner sep=3.2pt,
        font=\footnotesize,
        align=center
    },
    person/.style={draw=black!60, line width=0.55pt},
    boxn/.style={draw, rounded corners=2pt, line width=0.4pt, align=center, inner sep=3pt},
    structn/.style={boxn, draw=black!55, fill=black!4},
    tile/.style={rounded corners=1pt, line width=0.3pt, minimum size=4.6mm, inner sep=0pt},
    vis/.style={tile, draw=figVisible!75!black, fill=figVisible!15},
    hid/.style={tile, draw=figHidden!75!black, fill=figHidden!15,
                postaction={pattern=north east lines, pattern color=figHidden!75}},
    flow/.style={->, line width=0.5pt, draw=black!60},
    leak/.style={->, line width=0.7pt, dashed, draw=figHidden!85!black},
    chip/.style={rounded corners=2pt, fill=figChip, text=white, align=left, inner sep=4pt},
    panel/.style={rounded corners=4pt, line width=0.4pt,
                  minimum width=0.485\linewidth, minimum height=4.4cm, inner sep=0pt},
    sub/.style={font=\scriptsize},
    body/.style={align=left, font=\scriptsize},
]

\definecolor{fig1}{HTML}{A5CF83}
\node (1label) [font=\footnotesize,align=left,anchor=south west,text width=4.8cm,minimum height=32pt] {Multi-tenant databases store user data in shared physical structures and \textit{post-filter} data with a policy $\pred$ to provide isolation.};
\node (1circle) [font=\footnotesize,left=-3pt of 1label] {\circled{1}};
\begin{pgfonlayer}{background}
\node [rounded corners=2pt,fit=(1label) (1circle),fill=fig1!35] {};
\end{pgfonlayer}

\definecolor{fig2}{HTML}{F0E76F}
\node (2label) [font=\footnotesize,align=left,anchor=west,text width=3.6cm,minimum height=32pt] 
at ($(1label.east)+(24pt,0)$)
{Queries interact with non-visible data before $\pred$ is applied, resulting in side-channels.};
\node (2circle) [font=\footnotesize,left=-3pt of 2label] {\circled{2}};
\begin{pgfonlayer}{background}
\node [rounded corners=2pt,fit=(2label) (2circle),fill=fig2!35] {};
\end{pgfonlayer}

\definecolor{fig3}{HTML}{9A5B0C}
\node (3label) [font=\footnotesize,align=left,anchor=north west,text width=6.3cm,minimum height=10pt] 
at ($(2label.north east)+(24pt,-1.5pt)$)
{\textit{Rich queries}  amplify this to \textit{rich recovery}  in two cases:};
\node (3circle) [font=\footnotesize,left=-3pt of 3label] {\circled{3}};
\begin{pgfonlayer}{background}
\node [rounded corners=2pt,fit=(3label) (3circle),fill=fig3!15] {};
\end{pgfonlayer}

\definecolor{figChip}{HTML}{2C2C2A}
\node (rlsheader) [font=\footnotesize,align=left,anchor=north west,text width=6.79cm,minimum height=10pt] at ($(3circle.south west)+(0,-9pt)$) {\textbf{\S\ref{sec:rls}: Row-level security} (PostgreSQL) \\[1pt]
Due to query optimizer behavior, \\
RTT timing changes reveal if \\
queries match unseen rows.};
\node (rlschip) [font=\scriptsize,rounded corners=2pt, fill=figChip!90, text=white, align=center, inner sep=2pt, anchor=north, text width=6.5cm] at ($(rlsheader.south)+(0,-1pt)$)
    {full record reconstruction via ranges and conjunctions};
\begin{pgfonlayer}{background}
\node [rounded corners=2pt,fit=(rlsheader) (rlschip),fill=cyan!10] {};
\end{pgfonlayer}

\coordinate (lic) at ($(rlsheader)+(28pt,-9pt)$);
\begin{scope}[shift={(lic)}]
\draw[draw=black!55, line width=0.4pt] (-0.8,0) -- (0.75,0);
\draw[draw=black!65, line width=0.6pt] (-0.55,0) parabola bend (-0.35,0.30) (-0.15,0);
\draw[draw=blue!40!black, line width=0.8pt] (0,0) parabola bend (0.25,0.62) (0.5,0);
\node[font=\scriptsize, text=black, anchor=north] at (0,-0.0) {$\TimeOracle(\qry)$};
\node[font=\color{blue!40!black}\scriptsize,align=left] at (1.56,0.25)
{$\implies$ leaks that \\ \qquad\, a record\\ \quad\,\,\,\,\, matches $\qry$.};
\end{scope}

\node (dlsheader) [font=\footnotesize,align=left,anchor=north west,text width=6.79cm,minimum height=10pt] at ($(rlsheader.south west)+(0,-20pt)$) {\textbf{\S\ref{sec:dls}: Document-level security} (Elastic \& OpenSearch) \\[1pt]
Corpus-wide statistics and \\
prefix-expansion behavior \\
reveal terms from index.};
\node (dlschip) [font=\scriptsize,rounded corners=2pt, fill=figChip!90, text=white, align=center, inner sep=2pt, anchor=north, text width=6.5cm] at ($(dlsheader.south)+(0,-1pt)$)
    {approx. text recovery via term \& $n$-gram extraction};
\begin{pgfonlayer}{background}
\node [rounded corners=2pt,fit=(dlsheader) (dlschip),fill=magenta!10] {};
\end{pgfonlayer}

\coordinate (ric) at ($(dlsheader)+(25pt,-11pt)$);
\begin{scope}[shift={(ric)}]
\draw[draw=black!55, line width=0.4pt] (-0.8,0) -- (0.65,0);
\fill[black!50]        (-0.55,0) rectangle (-0.25,0.65);
\fill[black] (0.05,0)  rectangle (0.35,0.35);
\node[font=\scriptsize, anchor=north] at (-0.40,-0.0) {$\Ctrls$};
\node[font=\scriptsize, anchor=north] at (0.20,-0.0)  {$\Cands$};
\node[font=\color{red!60!black}\scriptsize,align=left] at (1.6,0.25) {$\implies$
  leaks that a \\ 
  \quad \,\,\,\,\,\,\,\, term with \\
  \quad \,\,\,\,\, prefix exists.};
\end{scope}

  \node[structn, minimum width=1cm, minimum height=0.95cm, anchor=south west] (adv) at ($(1label.south west)+(-12pt,-66pt)$)
        {Adversary\\[-1pt]
        \scriptsize (authorized\\[-2pt]
        \scriptsize user)};

  \node[structn, minimum width=2.75cm, minimum height=2.5cm, right=35pt of adv] (idx) {};
  \node[sub, anchor=north west, below right=1mm and 1mm of idx.north west] {Shared table $\tble$};
  \node[vis,anchor=north west] at ($(idx.north west)+(7pt,-16pt)$) (ta) {};
  \node[hid, right=1.2mm of ta]  (tb) {};
  \node[vis, right=1.2mm of tb]  (tc) {};
  \node[vis, right=1.2mm of tc] (td) {};
  \node[hid,below=1.3mm of ta] (taa) {};
  \node[hid,below=1.3mm of tb] (tbb) {};
  \node[vis,below=1.3mm of tc] (tcc) {};
  \node[hid,below=1.3mm of td] (tdd)  {};
  \node[vis,below=1.3mm of taa] (taaa) {};
  \node[hid,below=1.3mm of tbb] (tbbb) {};
  \node[hid,below=1.3mm of tcc] (tccc) {};
  \node[vis,below=1.3mm of tdd] (tddd)  {};
  \node[draw=black!55!green, dashed, rounded corners=2pt, line width=0.4pt,
        fit=(tc), inner sep=1.6pt] {};

  \node[draw=black!55!blue, dashed, rounded corners=2pt, line width=0.4pt,
        fit=(taaa)(tbbb), inner sep=1.6pt] {};

  \node[structn, minimum width=1cm, minimum height=0.95cm, right=18mm of idx] (pol)
        {Policy $\pred$\\[-1pt]\scriptsize (removes\\[-2pt]
        \scriptsize records)};
    
  \node[vis] at ($(idx.east)+(25pt,11pt)$) (res1) {};
  \coordinate (mpt) at ($(res1.north)!0.5!(res1.north)$);
  \node[anchor=south,color=black!55!green] at ($(mpt)+(0pt,0pt)$) {$\qry(\tble)$};

  \node[vis] at ($(idx.east)+(17pt,-23pt)$) (res1) {};
  \node[hid, right=1.2mm of res1] (res2) {};
  \coordinate (mpt) at ($(res1.north)!0.5!(res2.north)$);
  \node[anchor=south,color=black!55!blue] at ($(mpt)+(0pt,0pt)$) {$\qry'(\tble)$};

  \node[boxn, draw=figVisible!75!black, fill=figVisible!12,
        minimum width=1, minimum height=0.9cm, right=6mm of pol] (res)
        {Results\\[-1pt]\scriptsize (visible\\[-2pt]\scriptsize only)};

  \node[sub, anchor=south west] (legv) at ($(adv.north west) + (12pt,18pt)$) {visible};
  \node[vis, minimum size=2.6mm, left=1mm of legv] {};
  \node[sub, anchor=north west] (legh) at ([yshift=-0.3mm]legv.south west) {hidden};
  \node[hid, minimum size=2.6mm, left=1mm of legh] {};

  \draw[flow] (adv) -- (idx);
  \node[sub, anchor=south,color=black!55!green] at ($(adv.east)!0.5!(idx.west)$) {query $\qry$};
  \node[sub, anchor=south,color=black!55!blue] at ($(adv.east)!0.5!(idx.west)+(0,-15pt)$) {query $\qry'$};
  \draw[flow] (idx) -- (pol);
  \draw[flow] (pol) -- (res);

  \coordinate (lkm) at ($(adv.south)!0.5!(idx.south)$);
  \draw[leak] (res.south) to[out=-90, in=-50, looseness=0.47] (adv.south);
  \node[anchor=north, align=left, sub, text=figHidden!85!black] at ($(adv.south east)+(5pt,-26pt)$)
        {Timing / scoring behavior of\\
        {\color{black!55!green} $\qry$} and {\color{black!55!blue} $\qry'$} leads to side-channel.};

\end{tikzpicture}

\captionof{figure}{Overview of this paper.}
\label{fig:overview}

          \vspace{0.25\baselineskip}
       \end{center}
   }]
}

\definecolor{codegreen}{rgb}{0,0.6,0}
\definecolor{codegray}{rgb}{0.5,0.5,0.5}
\definecolor{codepurple}{rgb}{0.58,0,0.82}
\definecolor{backcolour}{rgb}{0.95,0.95,0.92}

\lstdefinestyle{mystyle}{
    backgroundcolor=\color{backcolour},   
    commentstyle=\color{codegreen},
    keywordstyle=\color{magenta},
    numberstyle=\tiny\color{codegray},
    stringstyle=\color{codepurple},
    basicstyle=\ttfamily\footnotesize,
    breakatwhitespace=false,         
    breaklines=true,                 
    captionpos=b,                    
    keepspaces=true,                 
    numbers=left,                    
    numbersep=5pt,                  
    showspaces=false,                
    showstringspaces=false,
    showtabs=false,                  
    tabsize=2
}

\newcommand{\HLColor}[2]{\sethlcolor{#1}\hl{#2}}

\allowdisplaybreaks 
\begin{document}

\date{}

\title{\Large \bf Plaintext Recovery Against Post-Filtering Access Control} 

\author{
{\rm Zachary Espiritu} \\
MongoDB Research  \and
{\rm David Cash} \\
University of Chicago
}

\maketitle

\begin{abstract} 

\textit{Fine-grained access control} (FGAC) mechanisms such as \textit{row-level
security} (RLS) and \textit{document-level security} (DLS) are widely deployed
in databases to restrict access to data stored in  physical indexing
structures shared by multiple users (e.g., in multi-tenant databases, or in
the implementation of least-privilege within an organization). 
FGAC implementations often use \emph{post-filtering} where untrusted queries
run over all data and private results are redacted afterwards. Prior work shows
this approach can lead to side-channels that enable attackers to test if a
chosen value exists in  unseen data.  While damaging, prior attacks do not
enable the efficient recovery of rich, high-entropy data like full records  or text documents.

We show  these side-channels are more damaging than previously thought.
Using \textit{rich query} interfaces (e.g.,  range, prefix, and
conjunctive predicates), we amplify existence leakage into 
reconstruction attacks.  We do this in two settings:
\begin{itemize}[itemsep=0pt,parsep=3pt,topsep=3pt,leftmargin=13pt]
\item \sethlcolor{cyan!10}\hl{\textbf{\,PostgreSQL (RLS timing).\,}} We exploit a timing side-channel
 and expressive SQL queries (e.g., ranges, conjunctions) to
enumerate unknown attribute values and, in turn, full records via binary search over large
domains.

\item \sethlcolor{magenta!10}\hl{\,\textbf{Elasticsearch/OpenSearch (DLS scoring).}\,} We exploit
scoring and prefix-expansion side-channels to recover indexed terms from 
documents. In some cases, we can extract 
$n$-grams in the corpus  to recover approximate  text.
\end{itemize}
Our results show that FGAC side-channels must be evaluated in the
presence of rich predicates, which can turn membership tests into scalable
reconstruction of high-entropy records.

\begin{tcolorbox}
This work was originally published by the USENIX Association in the \textit{Proceedings of the 35th USENIX Security Symposium} as \cite{originalpaper}. This full version contains expanded versions of \Cref{table:oracle-num-queries} and \Cref{figure:reconstruction-example} and the row-level security variable-length string recovery extension in Appendix~\ref{appendix:prefix-queries}.
\end{tcolorbox}
\end{abstract}

\section{Introduction}

\textit{Fine-grained access control} (FGAC) mechanisms such as \textit{row-level
security} (RLS)~\cite{noauthor_59_2025} in relational \textit{database management systems} (DBMS) and
\textit{document-level security}
(DLS)~\cite{noauthor_controlling_nodate,noauthor_document-level_2026} in
text-search DBMSes are widely deployed
to enforce isolation for users that are only permitted to access a subset of the
data. In such \textit{multi-tenant} DBMSes, data is co-located in shared
physical database indexing structures, but the FGAC access control system restricts
users to view or edit only a subset of the data satisfying a specific policy predicate.

Multi-tenancy is attractive for database administrators as it can standardize
and consolidate what would otherwise be distinct databases. This reduces
administrative and monetary overhead, but introduces additional risk with
having multiple tenants' data mingled within the same system. FGAC mitigates this risk by centralizing enforcement within the DBMS itself. This reduces the risk of application-level bugs and enables expressive query interfaces for authorized tenants. Because of these benefits, FGAC is used extensively in many industry applications \cite{harris-holt_friendly_2024,rousey_using_2023,pena_django_2024,prakash_enhancing_2025,brown_row_2025,goff_ensuring_2025,pokryvailo_shipping_2022,manouvrier_using_2022,chaudhary_protecting_2025}. %

The intended goal of FGAC is \textit{isolation}: users should not learn anything
about the database beyond what they are permitted to view. Towards this goal, FGAC
systems typically are implemented by \emph{post-filtering}: queries are run over
\emph{all} of the data, and then the system redacts private
data from the output before returning. 
Prior works showed  isolation  does not fully hold in post-filtering FGAC
systems by exploiting side-channels to construct
\textit{existence oracles} that allow an attacker to test if a chosen value
appears in the private data. This was done in two previous lines of work, but
their commonality has not been explicitly noted before this work.

For RLS, Dar, Hershcovitch, and Morrison \cite{dar_rls_2023}
and Rasin, Herbick, Lenard, Scope, and Wagner \cite{rasin_vulnerability_2024}
demonstrated that the execution time of SQL queries can reveal
whether a RLS-protected column contains an attacker-chosen value in some hidden row.  For DLS, Büttcher and Clarke \cite{buttcher2005security}
and Wang, Grubbs, Lu, Bindschaedler, Cash, and Ristenpart
\cite{wang_side-channel_2017} exploited full-text search systems that produce
ranked lists of results for queries.  They showed that corpus-level statistics
used to rank documents can reveal whether attacker-chosen keywords appear in the
private, DLS-protected corpus.

Initially, such leaks may appear limited in scope.   Testing for existence does
not immediately lead to the extraction of entire records. We could naively probe candidate values one by one, but for
large or high-entropy domains (e.g., SSNs, documents), linear probing is
impractical. Possibly for these reasons, prior
responses have largely dismissed these side-channels. For example, there is no evidence that GitHub patched their system
in response to the attack of Wang et al.~\cite{wang_side-channel_2017},  some RLS documentation only notes
the presence of a timing side
channel and dismisses the idea of fine-grained inference~\cite{bigquery,pg_rules_privs}\footnote{The latter
document~\cite{pg_rules_privs} notes that \emph{
  ``A malicious attacker might be
able to infer something about the amount of unseen data, or even gain some
information about the data distribution or most common values ...''}}, and a popular DLS system notes
possible scoring side-channels, but still states that DLS ``prevents users from
viewing restricted documents''~\cite{es_security_limitations}.
Dar et al. \cite{dar_rls_2023} speculated that timing side-channels could be
exploited to learn further information, but did not show how to do so
systematically or efficiently.

\heading{Contributions.} 
We give new attacks that go beyond presence detection and show FGAC
mechanisms can completely fail to keep data confidential.  Modern DBMSes allow users to issue range predicates, prefix queries,
conjunctions, and other expressive filters over indexed data,  and we show that
an attacker can adaptively issue these \textit{rich queries} to observe various 
side-channel signals about private data.  By stringing several of these queries together, our
attacks  reconstruct entire hidden rows or the approximate contents of hidden
documents and show that such side-channels are actually quite leaky in
light of the rich query interfaces that FGAC databases  support.

We study the security of post-filtering FGAC systems that support rich queries
in two  different contexts: relational databases and full-text search
systems. While based on a common conceptual issue, the mechanisms underlying the
attacks are completely different (e.g., timing versus scoring side-channels, exploiting different rich query types). Our case studies include some of the most
widely used software in both domains, namely PostgreSQL, Elasticsearch, and
OpenSearch.

Our first case-study concerns \sethlcolor{cyan!10}\hl{\textbf{\,RLS timing
side-channels\,}} \textbf{in PostgreSQL.} We show that the timing side-channel
identified in prior work can be efficiently leveraged far beyond presence
testing. By issuing range queries, an attacker can binary search over large
domains and recover exact attribute values. 
Then, by using conjunctions, the attacker can associate recovered values across
columns to reconstruct full rows. Our attacks apply to common SQL data types,
work with standard indexing configurations, and require no auxiliary knowledge
about the data distribution other than the schema. This improves prior work by
demonstrating the side-channel can be efficiently exploited for full record
recovery. 

Our second case-study examines \sethlcolor{magenta!10}\hl{\textbf{\,DLS scoring
side-channels\,}} \textbf{in Elasticsearch and OpenSearch.} We show 
scoring-based leakage in full-text search systems can be similarly amplified,
though via completely different means.  We identify a new \textit{prefix-expansion} side-channel in text completion queries
whose scores depend on corpus-wide statistics and construct oracles for prefix
and $n$-gram membership (depending on the specific indexing scheme).  Given
this, one can apply standard techniques that use $n$-grams to infer
text fragments from DLS-protected corpora.  This improves prior
attacks which could only test for exact tokens and could not
recover tokens letter-by-letter or detect tokens co-occurring in a 
document.

While we suggest mitigations that can partially address these attacks by
eliminating or reducing the side-channel, our combined 
results suggest that post-filtering is a fundamentally flawed approach to
implementing FGAC when rich queries are supported. Rich queries  depend
on the underlying data in complex ways (compared to simpler 
equality predicates). As we show, this empowers an attacker
who can abuse these post-filtering side-channels to exfiltrate significant information.

\heading{Organization.}
We describe the common threat model for all of our attacks in \Cref{sec:threatmodel}.
In Sections \ref{sec:rls}
and \ref{sec:dls}, we give self-contained case studies for RLS and DLS attacks.
 We  discuss related work in
 \Cref{sec:relatedwork} and conclude in \Cref{sec:conclusion}. 

\section{Abstract FGAC and Common Threat Model} 
\label{sec:threatmodel}

We first describe an abstract architecture that models all of the systems we
consider, and then describe the common threat model for our attacks.  We then
discuss some motivation scenarios where this threat model is appropriate.

\heading{Abstract \textit{post-filtering FGAC}.} The DBMS holds some data that we
denote $\datastore$, and supports the processing of queries $\qry$ against
$\datastore$. We denote the result of a query $\qry$ by $\qry(\datastore)$. Each user has an associated \textit{policy} predicate $\pred$  that determines what data they should be able to access.  When the user supplies a query $\qry$, the DBMS computes $\qry(\datastore)$ and then \emph{post-filters} this result by
computing and returning $\pred(\qry(\datastore))$.
Users may also be allowed to write and delete data to $\datastore$, subject
to some policy (e.g., only to records associated with their account).
The crucial feature of this architecture is that $\qry(\datastore)$, the
pre-filtered results, may contain data that the adversary is not authorized to
access, but system behavior may depend on $\qry(\datastore)$ in an
exploitable way.

\heading{Threat model.} The adversary is a legitimate user whose access is
mediated by the FGAC system holding $\datastore$ with some policy $\pred$. The
adversary aims to circumvent this mediation and recover as much unauthorized
data from $\datastore$ as possible.  
The adversary can issue queries $\qry$ of its choosing from some allowed set and
observe the post-filtered results, i.e.  $\pred(\qry(\datastore))$.  The
adversary might also be allowed to inject records into $\datastore$ under the allowed policy.
We assume the adversary has general knowledge of the type of data it is
attacking (e.g., the schema, or a subset of it, and general upper and lower bounds of attribute domains), but does not have distributional
knowledge about the contents.
Except for the  side-channels we analyze, we assume the FGAC system functions
correctly---the attacker cannot bypass policy enforcement or access raw storage,
and also cannot modify the records of other users.

Adversaries will abuse two types of side-channels: in \Cref{sec:rls},
they  measure the time needed to process and filter the query; in
\Cref{sec:dls}, they will examine the  response itself.

\heading{Motivating use-case: Direct access.} Many of the citations from the
introduction discuss giving an untrusted principal direct access to a database
where an organization stores data from several sources and attempts to restrict
internal analysts to only access subsets of the data. For example, these could
be data scientists working with health data  partitioned by
 project, and they could be allowed arbitrary SQL queries or text
searches.

An alternative version of this threat model arises when FGAC is used for
defense-in-depth. Here, several internal applications access a database.  To
limit the impact of a compromised application, least-privilege is implemented
using FGAC. When  an application is compromised, the adversary obtains
 query access similar to that of a trusted insider.

\heading{Motivating use-case: App-mediated access.} In the second 
use-case, a backend DBMS is only accessed by an intermediate app, such as a
webpage, which we assume is trusted. Users may be customers who register
accounts or employees of an organization. Data is restricted according to a
policy that will typically isolate users' data from other users.

Users  enter queries through the app interface which translates them into 
DBMS queries on behalf of the user. For example,
the GitHub deployment of Elasticsearch from~\cite{wang_side-channel_2017} allowed users (even unauthenticated) to
issue queries via an API and view the resulting documents. Another
example may include a hospital that allows non-technical users like doctors to
query data through a GUI; this system would restrict the doctors to only
viewing the data of their own patients. 

Such a system would allow the  attacker to  abuse only the queries triggered by
the app rather than arbitrary queries. In the case of a
relational DBMS, the reach of our attacks will correspond to which columns can
be queried and the allowed query types   (e.g., range, equality). For full-text search, we  assume the app permits the needed type of
query (e.g., exact-match or phrase prefix queries, as defined in~\Cref{sec:dls}).

\section{\colorbox{cyan!10}{RLS Timing Side-Channels}}
\label{sec:rls}

\setlength{\abovedisplayskip}{4.5pt}
\setlength{\belowdisplayskip}{4.5pt} 
\setlength{\abovedisplayshortskip}{2pt} 
\setlength{\belowdisplayshortskip}{2pt}  

We present our first case study of FGAC security:
timing attacks against \textit{row-level security}
(RLS) in relational databases, specifically PostgreSQL. We first
cover necessary background, explain why PostgreSQL's RLS implementation
enables  construction of a timing-based existence oracle, and then show how
expressive  predicates can amplify this oracle into end-to-end row
reconstruction.

\subsection{Background and Prior Work}\label{sec:rls-background}

\vspace{-3.5pt}
\heading{Instantiating the abstract FGAC model.} We target a PostgreSQL table  with several columns and rows. The table
will be protected by RLS with a policy that restricts the
adversary to only \texttt{SELECT} \textit{authorized} rows. The adversary aims to learn as much about \textit{unauthorized} rows as possible.
The adversary can issue \texttt{SELECT} queries from an expressive subset of
SQL, which are then run
against the  table and then post-filtered according to the policy.  The
adversary can observe the time elapsed between sending its query and
receiving the response (over a network). \Cref{fig:overview}
diagrams this architecture and basic mechanism for the side-channel we  exploit
and extend~\cite{dar_rls_2023}.

\heading{DBMS background.}
RLS is implemented within PostgreSQL's query engine. 
While post-filtering is not an explicit goal in PostgreSQL's RLS design, it arises naturally as a consequence of several 
other design goals.

Modern DBMSes (like PostgreSQL) have \textit{query optimizers} that determine how
to efficiently execute queries. Rather than directly executing a query as
written, the optimizer enumerates candidate physical query plans (ideally
leveraging precomputed physical indexes), estimates each plan's cost (e.g.,
based on the number of rows that are evaluated by each plan step), and then
executes the least costly plan.

In theory, the optimizer can choose to run the RLS filter before or
after the rest of the query. 
However, the execution order of the filter leads to some known folklore side-channels
against RLS \cite{sql_server_rls}---for instance, private values may determine whether or not
a query throws a division-by-zero error that is visible to the adversary.
To prevent this, PostgreSQL's optimizer provides the  \texttt{security\_barrier} feature which 
prevents evaluation of user-supplied portions of 
RLS-protected queries  against records \textit{before} the policy has been evaluated on
the record, effectively forcing \emph{pre}-filtering rather than \textit{post}-filtering.
This ensures that the user-defined  query portion only runs on RLS-allowed rows, but incurs significant performance penalties as it
restricts the optimizer's ability to reorder and combine query predicates. Forcing the policy to evaluate first often prohibits index usage, resulting in a sequential scan of the entire
table for every RLS-based query.

To mitigate this, PostgreSQL added a notion of  
\texttt{LEAKPROOF} operators that do not have any explicitly observable
side-effects, and allows for post-filtering on queries composed of only
\texttt{LEAKPROOF} operators.  Such operators include equality
(\texttt{=}) and comparison  (\texttt{<}, \texttt{>}, etc.).
If a predicate is entirely composed of \texttt{LEAKPROOF} operators, the
optimizer is allowed to reorder the evaluation of the predicate with respect to
the policy, so that the predicate is evaluated first. This allows
\texttt{LEAKPROOF} predicates to be pushed down into index access plans by the
query optimizer so that they may be evaluated before the RLS policy $\pred$ is
evaluated. This improves query performance, but simultaneously introduces a
timing side-channel.

\heading{The timing side-channel.} Dar et al. (DHM)
\cite{dar_rls_2023} identified a timing side-channel which arises from
PostgreSQL's aforementioned \texttt{LEAKPROOF} semantics.  For a query $\qry$
issued against a table $\tble$, let $\qtime(\qry,\tble)$ denote the wall-clock
execution time of evaluating $\qry$ on $\tble$, and write $\qry(\tble)$ for the
resulting output.  Then, let $\tble$ be a table protected by RLS with policy
predicate $\pred$; we will denote this by $\tble_\pred$. Semantically, the user
should observe the result $\qry(\tble_\pred) = \qry(\pred(\tble))$ or $\pred(\qry(\tble))$. If
the query optimizer chooses the latter---that is, it evaluates a predicate $\qry$ before the policy $\pred$, which
can happen if $\qry$ is composed entirely of \texttt{LEAKPROOF} operators \textit{and} if $\qry$ can be serviced by an index---then the time to evaluate $\qry(\tble_\pred)$ will be
\begin{align}\label{eq:time}
    \qtime(\qry, \tble_\pred) 
    & \approx \qtime(\qry,\tble) + \qtime(\pred,\qry(\tble)).
\end{align}
The key observation is that $\qtime(\pred,\qry(\tble))$ (the time to
apply $\pred$ to $\qry(\tble)$) depends on
$|\qry(\tble)|$, the size of the intermediate result \textit{before the RLS
policy $\pred$ has been applied}. Then, suppose the adversary knows a query predicate $\qry_0$ that 
returns zero rows. If the query optimizer evaluates $\qry_0$ before the
RLS filter, then $\qtime(\qry_0,\tble_\pred)$ is determined by
\Cref{eq:time}.
But because $\qry_0(\tble) = \emptyset$, it will
typically hold that $\qtime(\pred,\qry_0(\tble)) =
\qtime(\pred,\emptyset) \approx 0$, so
$\qtime(\qry_0, \tble_\pred)  \approx \qtime(\qry_0,\tble)$.

Now, let $\qry$ be another query for which the adversary would like to learn
$|\qry(\tble)|$. The $\qtime(\qry,
\tble_\pred)$ is again determined by \Cref{eq:time}.
In this case, putting everything together gives
\begin{align*}
  \qtime(\qry, \tble_\pred)& - \qtime(\qry_0, \tble_\pred) \\
    &\approx 
      (\qtime(\qry,\tble) + \qtime(\pred,\qry(\tble)))
      -
      \qtime(\qry_0,\tble) \\
    &=
      (\qtime(\qry,\tble) - 
      \qtime(\qry_0,\tble)) 
      + \qtime(\pred,\qry(\tble)).
\end{align*}
In practice, $(\qtime(\qry,\tble) - \qtime(\qry_0,\tble))$ tends to be
very small, even when $\qry$ and $\qry_0$ return a different number of rows.
As such, the main contributor to timing differences is
$\qtime(\pred,\qry(\tble))$ which itself  grows with $|\qry(\tble)|$---what the adversary wants to learn. 
Algorithm~\ref{alg:time_oracle} shows how DHM exploited this via a calibrated threshold test.
Given a probe $\qry$ and calibration queries $\qry_0,\qry_1$ with known
cardinality $0$ and $1$, it measures and takes the minimum runtime of each over $k$ trials (DHM set $k = 10$) and returns
\textsf{True} iff the measured time $t'$ for $\qry$ exceeds the midpoint
$(t_0+t_1)/2$.

\begin{figure}
\centering
\footnotesize
\begin{tikzpicture}
  \node[anchor=south west, inner sep=0pt] (hist) {\input{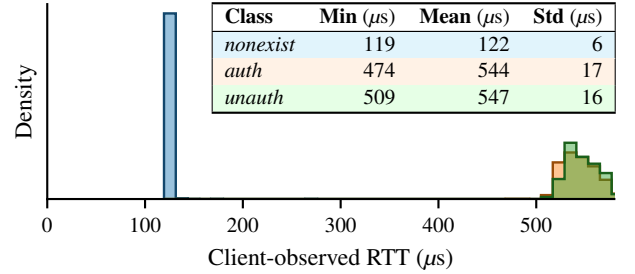}};
  \node[anchor=north east, inner sep=1pt, xshift=-2mm, yshift=-5.5pt]
       at (hist.north east) {%
       \footnotesize
\setlength{\aboverulesep}{0pt}
\setlength{\belowrulesep}{0pt}
       \setlength{\tabcolsep}{4.5pt}%
    \renewcommand{\arraystretch}{1.05}%
    \begin{tabular}{|l rrr|}
    \toprule
    \textbf{Class} & \textbf{Min} ($\mu$s) & \textbf{Mean} ($\mu$s) & \textbf{Std} ($\mu$s) \\
    \midrule
    \rowcolor{cyan!10}\textit{nonexist} & 119 & 122 & 6 \\
    \rowcolor{orange!10}\textit{auth} & 474 & 544 & 17 \\
    \rowcolor{green!10}\textit{unauth} & 509 & 547 & 16 \\
    \bottomrule
    \end{tabular}%
  };
\end{tikzpicture}
\vspace{-2.1em}
\caption{Equality query timing distributions from 1k trials in each class under a join-based RLS policy (see \Cref{sec:evaluation}) based on client observed RTT. \textbf{Takeaway:} \sethlcolor{orange!10}\hl{\,\emph{Auth}\,} and \sethlcolor{green!10}\hl{\,\emph{unauth}\,} queries are reliably longer than \sethlcolor{cyan!10}\hl{\,\emph{nonexist}\,} under join policies.}
\label{fig:existence-kde-join}
\end{figure}

In \Cref{fig:existence-kde-join}, we illustrate the timing side-channel on PostgreSQL; we defer experimental details to \Cref{sec:evaluation}. The \sethlcolor{cyan!10}\hl{\,\emph{nonexistent}\,} distribution corresponds to a probe predicate $\qry$ that matches no rows in the underlying table (e.g., $|\qry(\tble)|=0$). The \sethlcolor{orange!10}\hl{\,\emph{authorized}\,} distribution corresponds to a probe that matches a row the attacker is permitted to see, so the row is returned and the policy evaluation occurs on a non-empty intermediate result. The \sethlcolor{green!10}\hl{\,\emph{unauthorized}\,} distribution corresponds to a probe that matches a row prior to policy enforcement but is fully filtered by RLS; that is, $|\qry(\tble)|>0$ while $\qry(\tble_P)=\emptyset$. The gap between the nonexistent and unauthorized distributions is the signal exploited by our existence oracle: an increase in runtime indicates the predicate matched hidden rows.

\begin{algorithm}[t]
\footnotesize
   \SetKwInOut{Input}{Input}
   \SetKwInOut{Output}{Output} 
   \SetKwInOut{Parameter}{Parameter} 

   \caption{$\TimeOracle$}
   \label{alg:time_oracle}
   \Input{A query $\qry$ and calibration queries $(\qry_0, \qry_1)$ where  $|\qry_0(\tble)| = 0$ and  $|\qry_1(\tble)| = 1$; the number of trials $k$}
   \Output{\textsf{True} iff $|\qry(\tble)| > 0$}
   $t_0 \gets \min(\{\qtime(\qry_0,\tble_\pred) : i \in [k]\})$ \tcp*{Run each
   query $k$ times,}
   $t_1 \gets \min(\{\qtime(\qry_1,\tble_\pred) : i \in [k]\})$ \tcp*{then take
   the min of $k$~}
   $t' \gets \min(\{\qtime(\qry,\tble_\pred) : i \in [k]\})$\;
   \lIf{$t' > (t_1 + t_0)/2$}{Return \textsf{True}}
   \lElse{Return \textsf{False}}
\end{algorithm}

\heading{Existence oracle limitations.} DHM focused on equality-based
queries where values  are tested by the adversary one at a time by selecting rows with a given attribute
in a column. 
For example, by querying $\TimeOracle$ with all possible ages, the attacker can learn which ages are present in the table.
However, this limits  an attacker interested in rich data
exfiltration. First, the attack  does not
give a way to associate values across columns or reconstruct full rows. Second,
it requires the attacker to know candidate values in advance. For high-entropy attributes with large domains, the space of possibilities is too
large to test one by one.  Furthermore, DHM do not elaborate on their choice of $k=10$, and it is unclear if lower values of $k$ (which increases efficiency) provide acceptable accuracy.

The work of DHM briefly mentioned that the side-channel worked for range queries, but did not explore whether the side-channel exists for more complicated
queries that combine several conjunctions and range predicates, or how to
efficiently compose such queries for an attack. %
It is not \emph{a priori}
clear that the channel exists for arbitrarily complex queries, since query planner effects and the structure of the underlying indexes could 
affect the stability of timing-based oracles.  Line 1 of
Algorithm~\ref{alg:time_oracle} hints at this, since the constructed query
$\qry_0$ must be similar enough to $\qry$ in order to capture meaningful timing deltas.
Moreover, errors due to timing noise may compound, and it is not clear if it is possible or how
expensive it will be to handle this.   In this work, we
address these and other technical issues to extend the side-channel impact to full row
recovery.

\subsection{Our Timing-Based Attack}
\label{sec:reconstruction}

We now describe an end-to-end reconstruction attack which combines the timing
existence oracle with \textit{expressive queries} to efficiently reconstruct the
entire set of records in a RLS-protected table. Our techniques generalize
to all indexable PostgreSQL data types; here, we focus
on integers and strings. We demonstrate how a \textit{binary search}  with SQL range queries (\texttt{BETWEEN}) and set inclusion queries (\texttt{IN}) allows   efficient recovery of  indexed attributes. Then, we demonstrate how to use \textit{conjunctions} to reconstruct entire records. The attack has two phases: (i) enumerating the set of values
present in each attribute, and (ii) assembling  full tuples
using conjunctions. 

\heading{Notation.}
Let $\tble$ be a table with attributes $(\attr_1,\ldots,\attr_m)$.
For each attribute $\attr_i$, let $D_i$ denote its (known) domain and
$S_i \subseteq D_i$ be the unknown values that actually appear in $\tble$.
We will call $S_i$ the \emph{support of $\attr_i$}.

\subsubsection{Phase I: Attribute Enumeration}

\begin{figure}
\centering
\footnotesize
\begin{tikzpicture}
  \node[anchor=south west, inner sep=0pt] (hist) {\input{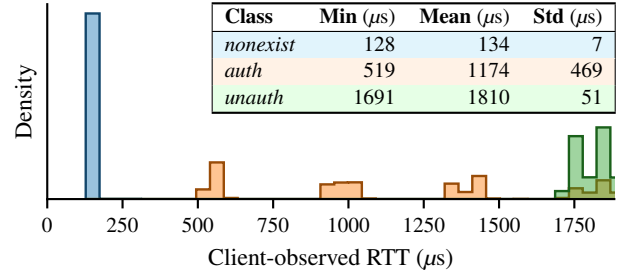}};
  \node[anchor=north east, inner sep=1pt, xshift=-2mm, yshift=-5.5pt]
       at (hist.north east) {%
       \footnotesize
\setlength{\aboverulesep}{0pt}
\setlength{\belowrulesep}{0pt}
       \setlength{\tabcolsep}{4.5pt}%
    \renewcommand{\arraystretch}{1.05}%
    \begin{tabular}{|l rrr|}
    \toprule
    \textbf{Class} & \textbf{Min} ($\mu$s) & \textbf{Mean} ($\mu$s) & \textbf{Std} ($\mu$s) \\
    \midrule
    \rowcolor{cyan!10}\textit{nonexist} & 128 & 134 & 7 \\
    \rowcolor{orange!10}\textit{auth} & 519 & 1174 & 469 \\
    \rowcolor{green!10}\textit{unauth} & 1691 & 1810 & 51 \\
    \bottomrule
    \end{tabular}%
  };
\end{tikzpicture}
\vspace{-2.1em}
\caption{Range  query timing distributions from 1k trials in each class  under a join-based  policy. On  \sethlcolor{orange!10}\hl{\,\textit{auth}\,} queries, the cardinality of $\qry(\tble)$ ranges from 1--4; the distinct peaks arise from different query cardinalities (which in turn check the policy different numbers of times). \textbf{Takeaway:} The timing side-channel is preserved under expressive queries.}
\label{fig:existence-range-kde-join}
\end{figure}

We first enumerate the support $S_i$ for each attribute independently.
To reduce the number of oracle queries needed to recover sensitive values, we use binary search over ordered attribute domains. Instead of probing each candidate value linearly, we issue range queries that test whether any matching records exist on one side of a split point. Since the B$^{+}$-tree indexes used in PostgreSQL can efficiently support range queries, the timing oracle still works provided that the \textit{shape} of the calibration queries is identical, allowing us to halve the remaining search space per query.
We validated that the timing oracle still works with expressive query types independently; microbenchmarks demonstrating this are in \Cref{fig:existence-range-kde-join}.\footnote{Range
queries can also be used to perform prefix queries to enumerate one character of a hidden string at a time, allowing us to recover strings of unknown length. (This works because strings are totally ordered on the B$^{+}$-tree indexes lexicographically under a deterministic collation that the attacker can order against.) We discuss this in more detail in  \Cref{appendix:prefix-queries}.}

\heading{Binary search subroutine.}
Given an attribute $\attr$ with ordered domain $D$, we recursively
partition the domain and query the oracle to test whether any values appear in a
given interval.  This procedure is formally outlined in
Algorithm~\ref{alg:enumerate}.  It uses
$\TimeOracle$ with a query $\qry$ of the form
\begin{lstlisting}[language=SQL,numbers=none,
    aboveskip=4pt,
    belowskip=3pt]
SELECT 1 FROM T WHERE atr >= L AND atr < R;
\end{lstlisting} %
If $\TimeOracle$ indicates  no record satisfies $\qry$, then Algorithm~\ref{alg:enumerate} stops and can conclude that there are no
records with attribute values in the range $[L, R)$; that is, $[L, R) \cap S_\attr =
\varnothing$. 
If $\TimeOracle$ indicates there is at least one record in the range $[L, R)$,
then the algorithm recursively checks $[L, M)$ and $[M, R)$ where $M = \lfloor (L +R)/2\rfloor$. This continues
until it reaches a unit interval ($L = R - 1$) where we can conclude $L \in S_\attr$.

\begin{algorithm}[t]
\footnotesize
\caption{\textsc{EnumerateAttribute}}
\label{alg:enumerate}
\KwIn{Attribute $\attr$ with ordered domain $[L,R)$}
\KwOut{Set $S_{\attr} \subseteq [L,R)$}
$S_{\attr} \gets \varnothing$\;
$\qry \gets \text{``$\attr \in [L,R)$''}$\;
\lIf{\textbf{not} $\TimeOracle(\qry)$}{
    \Return{$\varnothing$}
}
\lIf{$L = R - 1$}{
    \Return{$\{L\}$}
}
$M \gets \lfloor (L+R)/2 \rfloor$\;
$S_{\attr} \gets S_{\attr} \cup
    \textsc{EnumerateAttribute}(\attr, [L,M))$\;
$S_{\attr} \gets S_{\attr} \cup
    \textsc{EnumerateAttribute}(\attr, [M,R))$\;
\Return{$S_{\attr}$}
\end{algorithm}

\heading{Correctness.} Algorithm~\ref{alg:enumerate} explores an interval if and only if it contains
at least one value from $S_{\attr}$. Assuming a sound oracle for
non-emptiness, every value in $S_{\attr}$ is eventually isolated at a
unit interval and returned.

Of course,  the noise introduced by the oracle introduces one additional challenge: we need to ensure that the potential error introduced at each oracle test does not cause us to introduce false negatives and incorrectly eliminate significant portions of the domain higher in the traversal process. 
We mitigate false negatives by calibrating against a baseline predicate of
known cardinality one, and by exploiting the fact that larger ranges result in 
typically larger post-filtering costs. In Section~\ref{sec:evaluation}, we show
that this calibration strategy yields negligible false-negative rates in practice.

\heading{Complexity.}
Let $|D|$ be the size of the attribute domain and $|S|$ the number of values
present. The recursion visits at most $O(|S|\log(|D|))$ intervals, yielding
the same number of oracle queries. This improves over linear probing when
$|S| \ll |D|$, which is typical for high-entropy domains (e.g., SSNs).

\begin{algorithm}[t]
\footnotesize
\caption{\BinSrchNextVals}
\label{alg:nextvals}
\KwIn{Partial tuple $\vpart$, and set $V$ of possible values for next $\attr_i$}
\KwOut{Tuples of the form $\vpart$ with a value from $V$ appended}

$(v_1,\ldots,v_{i-1}) \gets \vpart$\;
\If{$|V| = 1$}{
  $\{v\} \gets V$\;
  $\qry \gets \text{``$\big ( \bigwedge_{j=1}^{i-1} (\attr_j = v_j) \big )
     \wedge (\attr_i = v)$''}$\;

   \lIf{$\TimeOracle(\qry)$}{\Return $\{(v_1,\ldots,v_{i-1},v)\}$}
   \lElse{\Return $\emptyset$}
}
\Else{
  Partition $V$ into two equal-sized sets $V_0,V_1$\;
  $\qry_0 \gets \text{``$\big ( \bigwedge_{j=1}^{i-1} (\attr_j = v_j) \big )
     \wedge (\attr_i \in V_0)$''}$\;
  $\qry_1 \gets \text{``$\big ( \bigwedge_{j=1}^{i-1} (\attr_j = v_j) \big )
     \wedge (\attr_i \in V_1)$''}$\;
   $C \gets \emptyset$\;
   \lIf{$\TimeOracle(\qry_0)$}{ $C \gets C\cup \BinSrchNextVals(\vpart,V_0)$}
   \lIf{$\TimeOracle(\qry_1)$}{ $C \gets C\cup \BinSrchNextVals(\vpart,V_1)$}
  \Return{$C$}
}
\end{algorithm}

\subsubsection{Phase II: Tuple Assembly via Conjunctions}
\label{section:tuple-assembly}

After recovering the approximate support $S_i$ of each 
$\attr_i$, we determine which combinations of values in $S_1 \times
\cdots \times S_n$ make up the records of the table. We do this by
incrementally constructing candidate tuples and using the timing oracle with
\textit{conjunctions} to perform cross-attribute existence checks. 
The algorithm initializes one partial tuple per value in $S_1$, then extends 
these tuples attribute-by-attribute by testing which values in $S_i$ co-occur 
with each surviving prefix.
A simple way to do this is to brute-force
every possible combination, but we do this  more efficiently using binary
search. Our strategy is given in Algorithms~\ref{alg:nextvals}
and~\ref{alg:tuples}, and we describe it here.

To recover values of attribute $\attr_2$ that appear with already-fixed
attribute $\attr_1$, we start by partitioning $S_2$ into two
sets $V_0,V_1$. We then determine if either of these sets contains a value
that co-occurs with $\attr_1$ by
running $\TimeOracle(\qry)$ 
with queries $\qry$ of the form 
\begin{lstlisting}[language=SQL,numbers=none,
    aboveskip=4pt,
    belowskip=3pt]
SELECT 1 FROM T WHERE atr_1=v_1 AND atr_2 IN V_b; 
\end{lstlisting} %
for $b=0,1$. When either of these queries returns true, we recursively
search the set $V_b$.  
Once we recover pairs of attributes  $(\attr_1,\attr_2)$, we  proceed to recover the third attribute. 

There are some practical subtleties in this phase as to how the ordering of attributes $S_1,\ldots,S_m$ in Algorithm~\ref{alg:tuples} affects the reconstruction quality and performance. We defer discussion of those details until our evaluation (\Cref{sec:evaluation}).

\begin{table*}[]
\footnotesize
    \centering
    \caption{Accuracy over 100k trials under CPU load generated by concurrent reads from other tenants. CPU load controlled to $\pm2\%$ of the target load. Wilson 99\% CIs \cite{Wilson01061927} in small text. \sethlcolor{yellow!0}\hl{\textbf{Takeaway:} For \textbf{join},  $k = \{1,2,3,4\}$ suffices for $>99.9\%$ accuracy at  CPU $\leq 95\%$. \textbf{inline} is significantly noisier. When load decreases, fewer queries suffice.}} 
\input{figures/table1_unified_full.pgf}
    \label{table:oracle-num-queries}
\end{table*}

\heading{Correctness.}
By construction, a partial tuple $(v_1,\ldots,v_i)$ is retained when at
least one row in $\tble$ matches the conjunction
$\bigwedge_{j=1}^i (\attr_j = v_j)$. Assuming perfect accuracy, if every row satisfies this
predicate for its own values, all rows are eventually reconstructed. 
If earlier stages of the reconstruction process have false positives, partial
tuples are discarded if the conjunction becomes unsatisfiable.

\begin{algorithm}[t]
\footnotesize
\caption{\ExtendTuples}
\label{alg:tuples}
\KwIn{Recovered supports $S_1,\ldots,S_m$} %
\KwOut{Set of reconstructed rows $C$}
$C \gets \{(v) : v \in S_1\}$\;
\For{$i \gets 2$ \KwTo $m$}{
    $C' \gets \emptyset$\;
    \For{$\langle v_1,\ldots,v_{i-1}\rangle \in C$}
    {
      $C' \gets C'\cup\BinSrchNextVals(\langle v_1,\ldots,v_{i-1}\rangle,S_{i})$
    }
    $C \gets C'$\;
}
\Return{$C$}
\end{algorithm}

\heading{Complexity.}
Let $n$ be the number of rows in $\tble$. 
The total number of calls to $\TimeOracle$ is bounded by
\[
  O\biggl(\sum_{i=2}^m |C_{i-1}| \log |S_{i}|\biggr)
 = O\biggl(n \sum_{i=2}^m \log |S_{i}|\biggr)
\]
where $|C_{i-1}|$ is the number of partial tuples found by depth $i-1$
(and $|C_{1}| =|S_1|$). 
When there are many tuples that share common prefixes, the latter formula
will be an overestimate. 

Combining Algorithms~\ref{alg:enumerate} and~\ref{alg:tuples} yields a complete
reconstruction attack. The total number of oracle queries is
\[
O\biggl(
\sum_{i=1}^m |S_i| \log{|D_i|} + 
n\sum_{i=2}^m \log |S_{i}|
\biggr),
\]
which is sublinear in the domain sizes and linear in the output size. In
Section~\ref{sec:evaluation}, we show that the timing oracle is strong enough for these bounds to translate into efficient  attacks.

\subsubsection{Optimizing \TimeOracle}\label{sec:rls-optimizations}

We finally identify an improved optimization for $\TimeOracle$ which leverages the fact that the information we learn from one $\TimeOracle$ call should help us in future oracle calls.
The first time we call $\TimeOracle$, we calculate a threshold $\Delta  = (t_1 -t_0)$ at the beginning of the attack. Then, on subsequent calls, we make a \textit{single} calibration query for $\qry_0$. We then decide membership based on if 
\[
\qtime(\qry,\tble_\pred) - 
\qtime(\qry_0,\tble_\pred) > \Delta / 2
\]
The intuition is that policy evaluation cost $\Delta$ should be roughly constant, so there is no need to recompute $\Delta$ after we have computed it the first time (assuming there was no unusual noise the first time). This reduces the number of queries per $\TimeOracle$ from $3$ to $2$, which in turn improves the robustness of the oracle since there are fewer opportunities for additional noise to be introduced.

\begin{table}[]
\centering
\footnotesize
\caption{Accuracy over 10k trials when attacker (\texttt{us-west3}) and DB (\texttt{us-east1}) are in separate regions ($\approx 3000$ km distance) on the \textbf{join} policy. \sethlcolor{yellow!0}\hl{\textbf{Takeaway:} Cross-region latency impacts accuracy but can be mitigated by increasing $k$.}}
\begin{tikzpicture}[x=2.0cm,y=0.37cm]
\definecolor{rlsOuter}{RGB}{34,34,34}
\definecolor{rlsText}{RGB}{17,17,17}
\definecolor{rlsHeaderText}{RGB}{34,34,34}
\definecolor{rlsWhite}{RGB}{255,255,255}
\node[font=\fontsize{6.5}{6.8}\selectfont, text=rlsHeaderText, inner sep=0pt, align=center] at (2.720,0.500) {\textbf{$k$ (Queries Per Probe Type)}};
\node[font=\fontsize{6.5}{6.8}\selectfont, text=rlsHeaderText, inner sep=0pt, align=center] at (0.230,-0.000) {\textbf{CPU}\\\textbf{Load}};
\node[font=\fontsize{6.5}{6.8}\selectfont, text=rlsHeaderText, inner sep=0pt, align=center] at (1.110,-0.000) {\textbf{Zone}};
\node[font=\fontsize{5.6}{6.0}\selectfont, text=rlsHeaderText, inner sep=0pt, align=center] at (2.000,-0.500) {1};
\node[font=\fontsize{5.6}{6.0}\selectfont, text=rlsHeaderText, inner sep=0pt, align=center] at (2.480,-0.500) {2};
\node[font=\fontsize{5.6}{6.0}\selectfont, text=rlsHeaderText, inner sep=0pt, align=center] at (2.960,-0.500) {4};
\node[font=\fontsize{5.6}{6.0}\selectfont, text=rlsHeaderText, inner sep=0pt, align=center] at (3.440,-0.500) {8};
\node[font=\fontsize{6.5}{7.0}\selectfont, text=rlsText, inner sep=0pt, align=center] at (0.230,-2.500) {Base};
\node[font=\fontsize{6.5}{7.0}\selectfont, text=rlsText, inner sep=0pt, align=center] at (0.230,-5.500) {50\%};
\node[font=\fontsize{6.5}{7.0}\selectfont, text=rlsText, inner sep=0pt, align=center] at (1.110,-1.500) {same zone};
\definecolor{rlsHeat0}{RGB}{0,104,55}
\path[fill=rlsHeat0, draw=none, line width=0pt] (1.760,-1.000) rectangle (2.240,-2.000);
\node[font=\fontsize{6.0}{6.5}\selectfont, text=rlsWhite, inner sep=0pt, align=center] at (2.000,-1.500) {100.0{\fontsize{4.3}{4.7}\selectfont\,$\pm$.00}};
\definecolor{rlsHeat1}{RGB}{0,104,55}
\path[fill=rlsHeat1, draw=none, line width=0pt] (2.240,-1.000) rectangle (2.720,-2.000);
\node[font=\fontsize{6.0}{6.5}\selectfont, text=rlsWhite, inner sep=0pt, align=center] at (2.480,-1.500) {100.0{\fontsize{4.3}{4.7}\selectfont\,$\pm$.00}};
\definecolor{rlsHeat2}{RGB}{0,104,55}
\path[fill=rlsHeat2, draw=none, line width=0pt] (2.720,-1.000) rectangle (3.200,-2.000);
\node[font=\fontsize{6.0}{6.5}\selectfont, text=rlsWhite, inner sep=0pt, align=center] at (2.960,-1.500) {100.0{\fontsize{4.3}{4.7}\selectfont\,$\pm$.00}};
\definecolor{rlsHeat3}{RGB}{0,104,55}
\path[fill=rlsHeat3, draw=none, line width=0pt] (3.200,-1.000) rectangle (3.680,-2.000);
\node[font=\fontsize{6.0}{6.5}\selectfont, text=rlsWhite, inner sep=0pt, align=center] at (3.440,-1.500) {100.0{\fontsize{4.3}{4.7}\selectfont\,$\pm$.00}};
\node[font=\fontsize{6.5}{7.0}\selectfont, text=rlsText, inner sep=0pt, align=center] at (1.110,-2.500) {\texttt{us-west3} $\leftrightarrow$ \texttt{us-east1}};
\definecolor{rlsHeat4}{RGB}{12,127,67}
\path[fill=rlsHeat4, draw=none, line width=0pt] (1.760,-2.000) rectangle (2.240,-3.000);
\node[font=\fontsize{6.0}{6.5}\selectfont, text=rlsWhite, inner sep=0pt, align=center] at (2.000,-2.500) {99.76{\fontsize{4.3}{4.7}\selectfont\,$\pm$.13}};
\definecolor{rlsHeat5}{RGB}{11,125,66}
\path[fill=rlsHeat5, draw=none, line width=0pt] (2.240,-2.000) rectangle (2.720,-3.000);
\node[font=\fontsize{6.0}{6.5}\selectfont, text=rlsWhite, inner sep=0pt, align=center] at (2.480,-2.500) {99.78{\fontsize{4.3}{4.7}\selectfont\,$\pm$.13}};
\definecolor{rlsHeat6}{RGB}{11,125,66}
\path[fill=rlsHeat6, draw=none, line width=0pt] (2.720,-2.000) rectangle (3.200,-3.000);
\node[font=\fontsize{6.0}{6.5}\selectfont, text=rlsWhite, inner sep=0pt, align=center] at (2.960,-2.500) {99.77{\fontsize{4.3}{4.7}\selectfont\,$\pm$.13}};
\definecolor{rlsHeat7}{RGB}{11,125,66}
\path[fill=rlsHeat7, draw=none, line width=0pt] (3.200,-2.000) rectangle (3.680,-3.000);
\node[font=\fontsize{6.0}{6.5}\selectfont, text=rlsWhite, inner sep=0pt, align=center] at (3.440,-2.500) {99.78{\fontsize{4.3}{4.7}\selectfont\,$\pm$.13}};
\node[font=\fontsize{6.5}{7.0}\bfseries\selectfont, text=rlsText, inner sep=0pt, align=center] at (1.110,-3.500) {$\Delta$};
\path[fill=rlsWhite, draw=none, line width=0pt] (1.760,-3.000) rectangle (2.240,-4.000);
\node[font=\fontsize{6.0}{6.5}\selectfont, text=rlsText, inner sep=0pt, align=center] at (2.000,-3.500) {$-$0.24{\fontsize{4.3}{4.7}\selectfont\,$\pm$.13}};
\path[fill=rlsWhite, draw=none, line width=0pt] (2.240,-3.000) rectangle (2.720,-4.000);
\node[font=\fontsize{6.0}{6.5}\selectfont, text=rlsText, inner sep=0pt, align=center] at (2.480,-3.500) {$-$0.22{\fontsize{4.3}{4.7}\selectfont\,$\pm$.13}};
\path[fill=rlsWhite, draw=none, line width=0pt] (2.720,-3.000) rectangle (3.200,-4.000);
\node[font=\fontsize{6.0}{6.5}\selectfont, text=rlsText, inner sep=0pt, align=center] at (2.960,-3.500) {$-$0.23{\fontsize{4.3}{4.7}\selectfont\,$\pm$.13}};
\path[fill=rlsWhite, draw=none, line width=0pt] (3.200,-3.000) rectangle (3.680,-4.000);
\node[font=\fontsize{6.0}{6.5}\selectfont, text=rlsText, inner sep=0pt, align=center] at (3.440,-3.500) {$-$0.22{\fontsize{4.3}{4.7}\selectfont\,$\pm$.13}};
\node[font=\fontsize{6.5}{7.0}\selectfont, text=rlsText, inner sep=0pt, align=center] at (1.110,-4.500) {same zone};
\definecolor{rlsHeat8}{RGB}{17,136,72}
\path[fill=rlsHeat8, draw=none, line width=0pt] (1.760,-4.000) rectangle (2.240,-5.000);
\node[font=\fontsize{6.0}{6.5}\selectfont, text=rlsWhite, inner sep=0pt, align=center] at (2.000,-4.500) {99.52{\fontsize{4.3}{4.7}\selectfont\,$\pm$.06}};
\definecolor{rlsHeat9}{RGB}{9,121,64}
\path[fill=rlsHeat9, draw=none, line width=0pt] (2.240,-4.000) rectangle (2.720,-5.000);
\node[font=\fontsize{6.0}{6.5}\selectfont, text=rlsWhite, inner sep=0pt, align=center] at (2.480,-4.500) {99.86{\fontsize{4.3}{4.7}\selectfont\,$\pm$.03}};
\definecolor{rlsHeat10}{RGB}{0,104,55}
\path[fill=rlsHeat10, draw=none, line width=0pt] (2.720,-4.000) rectangle (3.200,-5.000);
\node[font=\fontsize{6.0}{6.5}\selectfont, text=rlsWhite, inner sep=0pt, align=center] at (2.960,-4.500) {99.99{\fontsize{4.3}{4.7}\selectfont\,$\pm$.01}};
\definecolor{rlsHeat11}{RGB}{0,104,55}
\path[fill=rlsHeat11, draw=none, line width=0pt] (3.200,-4.000) rectangle (3.680,-5.000);
\node[font=\fontsize{6.0}{6.5}\selectfont, text=rlsWhite, inner sep=0pt, align=center] at (3.440,-4.500) {100.0{\fontsize{4.3}{4.7}\selectfont\,$\pm$.00}};
\node[font=\fontsize{6.5}{7.0}\selectfont, text=rlsText, inner sep=0pt, align=center] at (1.110,-5.500) {\texttt{us-west3} $\leftrightarrow$ \texttt{us-east1}};
\definecolor{rlsHeat12}{RGB}{107,191,100}
\path[fill=rlsHeat12, draw=none, line width=0pt] (1.760,-5.000) rectangle (2.240,-6.000);
\node[font=\fontsize{6.0}{6.5}\selectfont, text=rlsText, inner sep=0pt, align=center] at (2.000,-5.500) {95.45{\fontsize{4.3}{4.7}\selectfont\,$\pm$.54}};
\definecolor{rlsHeat13}{RGB}{27,153,80}
\path[fill=rlsHeat13, draw=none, line width=0pt] (2.240,-5.000) rectangle (2.720,-6.000);
\node[font=\fontsize{6.0}{6.5}\selectfont, text=rlsWhite, inner sep=0pt, align=center] at (2.480,-5.500) {98.90{\fontsize{4.3}{4.7}\selectfont\,$\pm$.27}};
\definecolor{rlsHeat14}{RGB}{19,140,74}
\path[fill=rlsHeat14, draw=none, line width=0pt] (2.720,-5.000) rectangle (3.200,-6.000);
\node[font=\fontsize{6.0}{6.5}\selectfont, text=rlsWhite, inner sep=0pt, align=center] at (2.960,-5.500) {99.40{\fontsize{4.3}{4.7}\selectfont\,$\pm$.20}};
\definecolor{rlsHeat15}{RGB}{12,127,67}
\path[fill=rlsHeat15, draw=none, line width=0pt] (3.200,-5.000) rectangle (3.680,-6.000);
\node[font=\fontsize{6.0}{6.5}\selectfont, text=rlsWhite, inner sep=0pt, align=center] at (3.440,-5.500) {99.74{\fontsize{4.3}{4.7}\selectfont\,$\pm$.14}};
\node[font=\fontsize{6.5}{7.0}\bfseries\selectfont, text=rlsText, inner sep=0pt, align=center] at (1.110,-6.500) {$\Delta$};
\path[fill=rlsWhite, draw=none, line width=0pt] (1.760,-6.000) rectangle (2.240,-7.000);
\node[font=\fontsize{6.0}{6.5}\selectfont, text=rlsText, inner sep=0pt, align=center] at (2.000,-6.500) {$-$4.07{\fontsize{4.3}{4.7}\selectfont\,$\pm$.54}};
\path[fill=rlsWhite, draw=none, line width=0pt] (2.240,-6.000) rectangle (2.720,-7.000);
\node[font=\fontsize{6.0}{6.5}\selectfont, text=rlsText, inner sep=0pt, align=center] at (2.480,-6.500) {$-$0.96{\fontsize{4.3}{4.7}\selectfont\,$\pm$.27}};
\path[fill=rlsWhite, draw=none, line width=0pt] (2.720,-6.000) rectangle (3.200,-7.000);
\node[font=\fontsize{6.0}{6.5}\selectfont, text=rlsText, inner sep=0pt, align=center] at (2.960,-6.500) {$-$0.59{\fontsize{4.3}{4.7}\selectfont\,$\pm$.20}};
\path[fill=rlsWhite, draw=none, line width=0pt] (3.200,-6.000) rectangle (3.680,-7.000);
\node[font=\fontsize{6.0}{6.5}\selectfont, text=rlsText, inner sep=0pt, align=center] at (3.440,-6.500) {$-$0.26{\fontsize{4.3}{4.7}\selectfont\,$\pm$.14}};
\draw[rlsOuter, line width=0.7pt] (0.000,1.000) -- (3.680,1.000);
\draw[rlsOuter, line width=0.4pt] (1.760,-0.000) -- (3.680,-0.000);
\draw[rlsOuter, line width=0.4pt] (0.000,-1.000) -- (3.680,-1.000);
\draw[rlsOuter, line width=0.5pt] (0.000,-4.000) -- (3.680,-4.000);
\draw[rlsOuter, line width=0.7pt] (0.000,-7.000) -- (3.680,-7.000);
\end{tikzpicture} 
    \label{table:oracle-cross-zone}
\end{table}

\subsection{End-to-End Evaluation}
\label{sec:evaluation}

We now  evaluate the rich-query-based timing attack.

\heading{Schema and data.} We replicate the example from Dar et al.~\cite{dar_rls_2023} for the use of FGAC in hospital electronic medical record (EMR) systems. We implement a hospital EMR schema containing a \texttt{patients} table with 1M rows (containing \texttt{id\_number}, \texttt{name}, \texttt{age}, \texttt{site\_id}, \texttt{zip\_code}, \texttt{ssn}) and a \texttt{doctors} table with 10k rows (mapping \texttt{user\_name} to \texttt{site\_id});  each patient attribute has its own B$^{+}$-tree index. 
Each doctor has their own clinic site (e.g., 10,000 sites total).

The full \texttt{patients} and \texttt{doctors} database occupies 185 MB on disk (79 MB heap, 37 MB \texttt{ssn} index, 11 MB \texttt{zip\_code} index, 7 MB \texttt{age} index, 51 MB other indexes), as determined by querying PostgreSQL's statistics after loading the dataset. With PostgreSQL's default shared buffer size of 128 MB and the operating system page cache, the data is resident in memory after it is queried the first time.\footnote{We confirm this using the \texttt{EXPLAIN BUFFERS} command on the \texttt{SELECT} queries against the database in our microbenchmarks.} This is a favorable condition for the attacker---workloads exceeding memory would add I/O variance that we do not evaluate, though  the adversary can partially mitigate  this by issuing each query an additional time to pull the necessary pages into the cache before actually making timing measurements.

\heading{Policy.} Often, a single EMR system is shared by multiple clinics who are subsidiaries of a parent organization. However, regulations require that doctors are only able to view patients who belong to their clinic.  This problem can be addressed by using a \textbf{join}-based PostgreSQL RLS policy which enforces that the \texttt{current\_user} can only see patients in the \texttt{patients} table who belong to the \texttt{current\_user}'s site.

\begin{lstlisting}[language=SQL,numbers=none,
    aboveskip=4pt,
    belowskip=3pt]
CREATE OR REPLACE FUNCTION site_policy_join(row_site BIGINT, curr_user TEXT)
RETURNS BOOLEAN AS $$ BEGIN
    RETURN (SELECT site_id FROM doctors WHERE 
            user_name = curr_user) = row_site; 
END; 
$$ LANGUAGE plpgsql;
CREATE POLICY doctor_read ON patients FOR SELECT USING site_policy_join(site_id, current_user);
\end{lstlisting}
We also tested an \textbf{inline}-based policy which replaces  \texttt{site\_policy\_join}  with a function that  accesses a session context variable \texttt{current\_setting(\textquotesingle{}app.site\_id\textquotesingle, true)} in lieu of a join. 
Both are representative of  real-world, deployed RLS policies~\cite{harris-holt_friendly_2024,rousey_using_2023,pena_django_2024,prakash_enhancing_2025,brown_row_2025,goff_ensuring_2025,pokryvailo_shipping_2022,manouvrier_using_2022,chaudhary_protecting_2025}. %

\heading{Setup.} We ran our experiments in Google Compute Engine (GCE). The server runs PostgreSQL 18.1 on Debian 12 on a  \texttt{c4-standard-16} host with 16 vCPUs (dedicated cores), 60 GB RAM, and a 200 GB local SSD with all of PostgreSQL's default settings. The attacker is located on a separate \texttt{c4-highmem-16} VM with 16 vCPUs and 128 GB RAM.  
The attacker remotely connects to the DBMS as a DBMS user  (corresponding to a doctor) with the ability to issue arbitrary \texttt{SELECT} queries on the \texttt{patients} table; their goal is to infer information about patients in a different site.
Latency is measured as client-side RTT with Python's \texttt{perf\_counter\_ns}. Unless otherwise stated, all VMs are located in the \texttt{us-central1} (Iowa, United States) region.

\begin{table*}[t]
\centering
\footnotesize
\setlength{\tabcolsep}{5pt}
\caption{Attribute reconstruction with $k = 1$ for oracle calls. Results averaged over $10$ repetitions with Student-$t$ 95\% CI in small text. Linear$^{\dagger}$ $\texttt{ssn}$ capped at $10^5$ probes; results are extrapolated to the full $10^9$ domain. We omit parallelism experiments from \texttt{age} for space; trends are similar to \texttt{zip\_code}.
\textbf{Takeaway:} When $\lvert D\rvert \gg \lvert S\rvert$ (e.g., \texttt{ssn}), binary search yields large speedups
over linear probing. When $\lvert D\rvert \approx \lvert S\rvert$ (e.g., \texttt{zip\_code}, \texttt{age}), linear probing is already practical and binary search slows the attack. Parallelism speeds up the attack, but too many threads decreases performance due to increased (self-imposed) DB load.}
\label{table:rls-reconstruction}
\begin{tabular}{r r r l l c r@{}l r@{}l r r r}
\toprule
\textbf{Attribute} & \textbf{\boldmath$|D_{\texttt{atr}}|$} & \textbf{\boldmath$|S_{\texttt{atr}}|$} & \textbf{Type} & \textbf{Strategy} & \textbf{Workers} & \multicolumn{2}{c}{\textbf{Time} (m)} & \multicolumn{2}{c}{\textbf{Queries}}& \multicolumn{1}{c}{\textbf{Recall}} & \multicolumn{1}{c}{\textbf{Precision}} & \textbf{Speedup} \\
\midrule
\multirow{6}{*}{\texttt{ssn}} & \multirow{6}{*}{$10^{9}$} & \multirow{6}{*}{$10^{6}$} & \multirow{6}{*}{\texttt{VARCHAR}} & Linear$^{\dagger}$ & 1 & 4187.6 & \,\scalebox{0.7}{$\pm$\,\makebox[\widthof{1138}][r]{1138}.9} & 2{,}000{,}000{,}006 & \,\scalebox{0.7}{$\pm$\,\makebox[\widthof{3{,}296{,}512}][r]{0}} & \cellcolor[RGB]{0,104,55}\textcolor{white}{100.00\makebox[\widthof{\,\scalebox{0.7}{$\pm$\,\makebox[\widthof{9}][r]{9}.50}}][l]{\,\scalebox{0.7}{$\pm$\,\makebox[\widthof{9}][r]{}.00}}} & \cellcolor[RGB]{0,104,55}\textcolor{white}{100.00\makebox[\widthof{\,\scalebox{0.7}{$\pm$\,\makebox[\widthof{9}][r]{9}.50}}][l]{\,\scalebox{0.7}{$\pm$\,\makebox[\widthof{9}][r]{}.00}}} & 1.0$\times$ \\
& & & & Binary & 1 & 150.4 & \,\scalebox{0.7}{$\pm$\,\makebox[\widthof{1138}][r]{13}.7} & 40{,}427{,}756 & \,\scalebox{0.7}{$\pm$\,\makebox[\widthof{3{,}296{,}512}][r]{2{,}537}} & \cellcolor[RGB]{0,104,55}\textcolor{white}{99.99\makebox[\widthof{\,\scalebox{0.7}{$\pm$\,\makebox[\widthof{9}][r]{9}.50}}][l]{\,\scalebox{0.7}{$\pm$\,\makebox[\widthof{9}][r]{}.01}}} & \cellcolor[RGB]{0,104,55}\textcolor{white}{99.99\makebox[\widthof{\,\scalebox{0.7}{$\pm$\,\makebox[\widthof{9}][r]{9}.50}}][l]{\,\scalebox{0.7}{$\pm$\,\makebox[\widthof{9}][r]{}.00}}} & 27.8$\times$ \\
& & & & Binary & 2 & 82.2 & \,\scalebox{0.7}{$\pm$\,\makebox[\widthof{1138}][r]{8}.2} & 40{,}428{,}226 & \,\scalebox{0.7}{$\pm$\,\makebox[\widthof{3{,}296{,}512}][r]{1{,}812}} & \cellcolor[RGB]{0,104,55}\textcolor{white}{99.99\makebox[\widthof{\,\scalebox{0.7}{$\pm$\,\makebox[\widthof{9}][r]{9}.50}}][l]{\,\scalebox{0.7}{$\pm$\,\makebox[\widthof{9}][r]{}.01}}} & \cellcolor[RGB]{0,104,55}\textcolor{white}{99.99\makebox[\widthof{\,\scalebox{0.7}{$\pm$\,\makebox[\widthof{9}][r]{9}.50}}][l]{\,\scalebox{0.7}{$\pm$\,\makebox[\widthof{9}][r]{}.00}}} & 50.9$\times$ \\
& & & & Binary & 4 & 42.0 & \,\scalebox{0.7}{$\pm$\,\makebox[\widthof{1138}][r]{3}.2} & 40{,}333{,}805 & \,\scalebox{0.7}{$\pm$\,\makebox[\widthof{3{,}296{,}512}][r]{79{,}723}} & \cellcolor[RGB]{102,189,99}99.49\makebox[\widthof{\,\scalebox{0.7}{$\pm$\,\makebox[\widthof{9}][r]{9}.50}}][l]{\,\scalebox{0.7}{$\pm$\,\makebox[\widthof{9}][r]{}.41}} & \cellcolor[RGB]{30,154,81}\textcolor{white}{99.93\makebox[\widthof{\,\scalebox{0.7}{$\pm$\,\makebox[\widthof{9}][r]{9}.50}}][l]{\,\scalebox{0.7}{$\pm$\,\makebox[\widthof{9}][r]{}.07}}} & 99.8$\times$ \\
& & & & Binary & 8 & 25.4 & \,\scalebox{0.7}{$\pm$\,\makebox[\widthof{1138}][r]{3}.0} & 28{,}753{,}796 & \,\scalebox{0.7}{$\pm$\,\makebox[\widthof{3{,}296{,}512}][r]{3{,}296{,}512}} & \cellcolor[RGB]{243,107,66}\textcolor{white}{51.85\makebox[\widthof{\,\scalebox{0.7}{$\pm$\,\makebox[\widthof{9}][r]{9}.50}}][l]{\,\scalebox{0.7}{$\pm$\,\makebox[\widthof{9}][r]{9}.50}}} & \cellcolor[RGB]{235,247,163}93.20\makebox[\widthof{\,\scalebox{0.7}{$\pm$\,\makebox[\widthof{9}][r]{9}.50}}][l]{\,\scalebox{0.7}{$\pm$\,\makebox[\widthof{9}][r]{1}.70}} & 165.1$\times$ \\
& & & & Binary & 16 & 11.3 & \,\scalebox{0.7}{$\pm$\,\makebox[\widthof{1138}][r]{}.9} & 11{,}647{,}270 & \,\scalebox{0.7}{$\pm$\,\makebox[\widthof{3{,}296{,}512}][r]{1{,}162{,}906}} & \cellcolor[RGB]{177,11,38}\textcolor{white}{7.77\makebox[\widthof{\,\scalebox{0.7}{$\pm$\,\makebox[\widthof{9}][r]{9}.50}}][l]{\,\scalebox{0.7}{$\pm$\,\makebox[\widthof{9}][r]{1}.24}}} & \cellcolor[RGB]{253,179,101}71.17\makebox[\widthof{\,\scalebox{0.7}{$\pm$\,\makebox[\widthof{9}][r]{9}.50}}][l]{\,\scalebox{0.7}{$\pm$\,\makebox[\widthof{9}][r]{2}.83}} & 371.1$\times$ \\
\midrule
\multirow{6}{*}{\texttt{zip\_code}} & \multirow{6}{*}{$10^{5}$} & \multirow{6}{*}{99{,}995} & \multirow{6}{*}{\texttt{VARCHAR}} & Linear & 1 & 3.3 & \,\scalebox{0.7}{$\pm$\,\makebox[\widthof{1138}][r]{}.1} & 200{,}006 & \,\scalebox{0.7}{$\pm$\,\makebox[\widthof{3{,}296{,}512}][r]{0}} & \cellcolor[RGB]{0,104,55}\textcolor{white}{100.00\makebox[\widthof{\,\scalebox{0.7}{$\pm$\,\makebox[\widthof{9}][r]{9}.50}}][l]{\,\scalebox{0.7}{$\pm$\,\makebox[\widthof{9}][r]{}.00}}} & \cellcolor[RGB]{0,104,55}\textcolor{white}{100.00\makebox[\widthof{\,\scalebox{0.7}{$\pm$\,\makebox[\widthof{9}][r]{9}.50}}][l]{\,\scalebox{0.7}{$\pm$\,\makebox[\widthof{9}][r]{}.00}}} & 1.0$\times$ \\
& & & & Binary & 1 & 7.0 & \,\scalebox{0.7}{$\pm$\,\makebox[\widthof{1138}][r]{}.2} & 400{,}004 & \,\scalebox{0.7}{$\pm$\,\makebox[\widthof{3{,}296{,}512}][r]{1}} & \cellcolor[RGB]{0,104,55}\textcolor{white}{99.99\makebox[\widthof{\,\scalebox{0.7}{$\pm$\,\makebox[\widthof{9}][r]{9}.50}}][l]{\,\scalebox{0.7}{$\pm$\,\makebox[\widthof{9}][r]{}.00}}} & \cellcolor[RGB]{0,104,55}\textcolor{white}{100.00\makebox[\widthof{\,\scalebox{0.7}{$\pm$\,\makebox[\widthof{9}][r]{9}.50}}][l]{\,\scalebox{0.7}{$\pm$\,\makebox[\widthof{9}][r]{}.00}}} & 0.5$\times$ \\
& & & & Binary & 2 & 3.5 & \,\scalebox{0.7}{$\pm$\,\makebox[\widthof{1138}][r]{}.1} & 400{,}000 & \,\scalebox{0.7}{$\pm$\,\makebox[\widthof{3{,}296{,}512}][r]{2}} & \cellcolor[RGB]{0,104,55}\textcolor{white}{99.99\makebox[\widthof{\,\scalebox{0.7}{$\pm$\,\makebox[\widthof{9}][r]{9}.50}}][l]{\,\scalebox{0.7}{$\pm$\,\makebox[\widthof{9}][r]{}.00}}} & \cellcolor[RGB]{0,104,55}\textcolor{white}{100.00\makebox[\widthof{\,\scalebox{0.7}{$\pm$\,\makebox[\widthof{9}][r]{9}.50}}][l]{\,\scalebox{0.7}{$\pm$\,\makebox[\widthof{9}][r]{}.00}}} & 0.9$\times$ \\
& & & & Binary & 4 & 1.8 & \,\scalebox{0.7}{$\pm$\,\makebox[\widthof{1138}][r]{}.1} & 394{,}991 & \,\scalebox{0.7}{$\pm$\,\makebox[\widthof{3{,}296{,}512}][r]{7{,}540}} & \cellcolor[RGB]{145,208,104}98.74\makebox[\widthof{\,\scalebox{0.7}{$\pm$\,\makebox[\widthof{9}][r]{9}.50}}][l]{\,\scalebox{0.7}{$\pm$\,\makebox[\widthof{9}][r]{1}.26}} & \cellcolor[RGB]{0,104,55}\textcolor{white}{100.00\makebox[\widthof{\,\scalebox{0.7}{$\pm$\,\makebox[\widthof{9}][r]{9}.50}}][l]{\,\scalebox{0.7}{$\pm$\,\makebox[\widthof{9}][r]{}.00}}} & 1.8$\times$ \\
& & & & Binary & 8 & 0.9 & \,\scalebox{0.7}{$\pm$\,\makebox[\widthof{1138}][r]{}.0} & 396{,}983 & \,\scalebox{0.7}{$\pm$\,\makebox[\widthof{3{,}296{,}512}][r]{2{,}951}} & \cellcolor[RGB]{122,198,101}99.20\makebox[\widthof{\,\scalebox{0.7}{$\pm$\,\makebox[\widthof{9}][r]{9}.50}}][l]{\,\scalebox{0.7}{$\pm$\,\makebox[\widthof{9}][r]{}.74}} & \cellcolor[RGB]{0,104,55}\textcolor{white}{100.00\makebox[\widthof{\,\scalebox{0.7}{$\pm$\,\makebox[\widthof{9}][r]{9}.50}}][l]{\,\scalebox{0.7}{$\pm$\,\makebox[\widthof{9}][r]{}.00}}} & 3.6$\times$ \\
& & & & Binary & 16 & 0.7 & \,\scalebox{0.7}{$\pm$\,\makebox[\widthof{1138}][r]{}.0} & 391{,}276 & \,\scalebox{0.7}{$\pm$\,\makebox[\widthof{3{,}296{,}512}][r]{3{,}600}} & \cellcolor[RGB]{179,223,114}97.59\makebox[\widthof{\,\scalebox{0.7}{$\pm$\,\makebox[\widthof{9}][r]{9}.50}}][l]{\,\scalebox{0.7}{$\pm$\,\makebox[\widthof{9}][r]{}.90}} & \cellcolor[RGB]{0,104,55}\textcolor{white}{100.00\makebox[\widthof{\,\scalebox{0.7}{$\pm$\,\makebox[\widthof{9}][r]{9}.50}}][l]{\,\scalebox{0.7}{$\pm$\,\makebox[\widthof{9}][r]{}.00}}} & 4.8$\times$ \\
\midrule
\multirow{2}{*}{\texttt{age}} & \multirow{2}{*}{120} & \multirow{2}{*}{120} & \multirow{2}{*}{\texttt{INTEGER}} & Linear & 1 & $<$0.1 &  & 246 & \,\scalebox{0.7}{$\pm$\,\makebox[\widthof{3{,}296{,}512}][r]{0}} & \cellcolor[RGB]{0,104,55}\textcolor{white}{100.00\makebox[\widthof{\,\scalebox{0.7}{$\pm$\,\makebox[\widthof{9}][r]{9}.50}}][l]{\,\scalebox{0.7}{$\pm$\,\makebox[\widthof{9}][r]{}.00}}} & \cellcolor[RGB]{0,104,55}\textcolor{white}{100.00\makebox[\widthof{\,\scalebox{0.7}{$\pm$\,\makebox[\widthof{9}][r]{9}.50}}][l]{\,\scalebox{0.7}{$\pm$\,\makebox[\widthof{9}][r]{}.00}}} & 1.0$\times$ \\
& & & & Binary & 1 & $<$0.1 &  & 484 & \,\scalebox{0.7}{$\pm$\,\makebox[\widthof{3{,}296{,}512}][r]{0}} & \cellcolor[RGB]{0,104,55}\textcolor{white}{100.00\makebox[\widthof{\,\scalebox{0.7}{$\pm$\,\makebox[\widthof{9}][r]{9}.50}}][l]{\,\scalebox{0.7}{$\pm$\,\makebox[\widthof{9}][r]{}.00}}} & \cellcolor[RGB]{0,104,55}\textcolor{white}{100.00\makebox[\widthof{\,\scalebox{0.7}{$\pm$\,\makebox[\widthof{9}][r]{9}.50}}][l]{\,\scalebox{0.7}{$\pm$\,\makebox[\widthof{9}][r]{}.00}}} & 0.5$\times$ \\
\bottomrule
\end{tabular}
\end{table*}

\subsubsection{Oracle Strength and Calibration}
\label{sec:oracle-strength}

An important detail  not explored in prior work \cite{dar_rls_2023,rasin_vulnerability_2024} is the number of queries necessary to perform the attack. This impacts the attack's efficiency and correctness.
To determine the number of repetitions $k$ per query type needed to instantiate the existence oracle in Algorithm~\ref{alg:time_oracle}, we compare the execution time of a
candidate predicate against a baseline predicate of identical syntactic
shape that is known to match zero rows.

We also evaluated how this changes based on noise introduced by other tenant reads on the table using a separate noise VM (\texttt{c4-standard-96} on GCE) that runs up to 96 concurrent doctor clients that issue a random mix of authorized/unauthorized/nonexistent \texttt{SELECT} queries to the database. We   target a specific CPU load by adjusting the noise VM's query rate depending on the DBMS's CPU load, as monitored via \texttt{vmstat}~\cite{vmstat}.
\Cref{table:oracle-num-queries} shows increased CPU load degrades  attack quality but can be overcome by issuing more queries.

We emphasize the \# of queries required for 99.9\% accuracy as the adaptiveness of our attacks from \Cref{sec:reconstruction} gain efficiency in exchange for the possibility of compounding oracle errors over time. For example, binary searching over the \texttt{ssn} attribute with $|D|=10^9$ requires $\log(10^9) \approx 30$ successful oracle calls to reach an attribute. Roughly speaking, with $99.9\%$ oracle accuracy, we expect $0.999^{30} \approx 97\%$ accuracy for SSN traversal. With $99.0\%$ oracle accuracy, we expect $0.99^{30} \approx 74\%$ accuracy for a given SSN.\footnote{As a remark, this assumes that the accuracy of oracle calls is independent, which is not strictly true due to cache effects.}

\heading{Physical distance.} We also evaluated how  physical distance from the target database affects the attack in \Cref{table:oracle-cross-zone} by placing the attacker in \texttt{us-west3} (Salt Lake City, USA) and the database in \texttt{us-east1} (South Carolina, USA). As shown, this results in a small loss of accuracy, but one that can be overcome simply by issuing more queries. Repeating the CPU load experiment in this setting shows that additional CPU load introduced by other tenants\footnote{In cross-zone experiments, we place the noise VM in the same region as the database---placing it in the farther region prevents the noise VM from reaching the required CPU load on the database due to the increased RTT.} similarly degrades accuracy. We remark our measurements use direct client-to-server connections; intermediaries such as load balancers could introduce additional variance we do not model.

\heading{Transparent data encryption.}
We remark that \textit{transparent data
encryption} (TDE) is frequently used in a DBMS to encrypt files
before writing them to disk.
PostgreSQL  does not have native
TDE but suggests LUKS~\cite{fruhwirth_luks} for index encryption on Linux hosts.
TDE does not  change the timing differences we exploit; after a page is decrypted and loaded into memory,  queries do not incur any I/O overhead and thus LUKS adds no noise to measurements. We validated  this  by reproducing the microbenchmarks from
\Cref{fig:existence-kde-join} and \Cref{fig:existence-range-kde-join} on a database  with LUKS2 enabled on the database directories. Since the results are essentially identical, we omit the figures.

\subsubsection{Record Enumeration}
\label{sec:attr-enum}

We now  evaluate the attribute enumeration procedure from  Algorithm~\ref{alg:enumerate} with the optimized $\TimeOracle$ from \Cref{sec:rls-optimizations} with $k = 1$ repetitions per query type. 
We report on the \emph{number of queries} required to
recover each attribute; this metric  reflects the stealth of the  attack
(fewer queries are harder to detect), wall clock time, and the \textit{recall} (TP / (TP + FN)) and \textit{precision} (TP / (TP + FP)) for each attribute. We also compare against the linear probing technique from DHM, but for fairness we also implement it with the optimized $\TimeOracle$ from \Cref{sec:rls-optimizations} and also set $k = 1$. 

\Cref{table:rls-reconstruction} shows our results. 
Binary search
recovers most values with minimal additional error; in contrast, linear probing~\cite{dar_rls_2023} requires orders of magnitude more queries and
is infeasible for high-entropy domains. We remark that the DHM work specified $k=10$ trials per query type, so their original attack would require $10\times$ more queries. Their original $\TimeOracle$ would also require an additional query per oracle call, which would require   $1.5\times$ more queries on top of that.

\heading{Parallelism.} We also parallelize the enumeration procedures in Section~\ref{sec:reconstruction}. In theory, as our oracles are read-only, parallel execution does not require coordination
beyond distributing intervals/partial tuples and aggregating oracle outcomes. In Algorithm~\ref{alg:enumerate} and Algorithm~\ref{alg:nextvals}, each oracle probe is a read-only query and we can concurrently execute recursion on disjoint
intervals. Concretely, we assign each worker a distinct child interval one level below the root, which lets $W$ workers explore $W$ disjoint subtrees independently.

As shown in \Cref{table:rls-reconstruction},  moving from $1$ to $W$ worker threads results in  close to a $W\times$
reduction in runtime---though past $W > 4$, the accuracy degrades as workers contend for the same database-side resources like the page cache and 
bandwidth.

\heading{Tuple recovery.}
We evaluate full row reconstruction using conjunctions as
described in \Cref{section:tuple-assembly}. We sample 100k rows from the dataset and use the support of the attributes across the sampled rows as the input to Algorithm~\ref{alg:tuples}. 
A subtlety in this phase is that conjunctive predicates' interaction with the DBMS's indexes 
  can introduce additional variance.
When multiple single-column indexes are available, PostgreSQL commonly evaluates a conjunction
by performing separate index scans and then intersecting their result sets (via \textit{bitmap
index scans}).
Such plans' execution time depends not only on the cardinality of the final result, but
also on the sizes of intermediate index results.
Even when the \emph{final} conjunction is empty, a probe where one conjunct
matches many rows and another matches none can be noticeably slower than a probe where both
conjuncts match none, because the engine still performs work to scan and combine intermediate
bitmaps.

As a result, the timing oracle is most stable when conjunctions are constructed
by starting from the most selective predicates. Starting from a selective attribute
(e.g., a unique or near-unique identifier) ensures that all subsequent
conjunctions operate over small intermediate result sets, strengthening the
timing signal and limiting extraneous intermediate tuple checks.\footnote{Starting from highly selective attributes also lets us recover \emph{unindexed} columns: as long as the unindexed column’s predicate is \texttt{LEAKPROOF}, PostgreSQL evaluates it in the final filter and short-circuits as soon as an earlier conjunct fails. When the RLS policy is evaluated last, this short-circuiting keeps the additional work (and thus the timing variance) concentrated in the policy evaluation, so the adversary can still  capture the necessary signal.}
This ordering is therefore essential both to control the number of oracle queries
and to maintain a reliable distinction between empty and non-empty predicates.
Accordingly, we build tuples by adding attributes in order of \emph{decreasing selectivity}
(i.e., from highest to lowest cardinality support): we iterate over attributes
$S_1,\ldots,S_m$ such that $|S_1| > \cdots > |S_m|$.

The results are in \Cref{table:tuplext-reconstruction}.\footnote{Tuple assembly is also parallelizable in principle as candidate extensions for a prefix are logically independent. Due to space restrictions and to reduce implementation complexity, we leave tuple-assembly results single-threaded.}  
As expected, total query counts scale  with the number of rows
reconstructed and remain within a small constant factor of the bound
derived in \Cref{section:tuple-assembly}. The results also confirm our intuition, which is that the most efficient order is in decreasing selectivity.

\begin{table}[t]
\centering
\footnotesize
\setlength{\tabcolsep}{4pt}
\caption{Single-threaded tuple  reconstruction on a 100k row target set. Student-$t$ 95\% CI over $3$ repetitions in small text. \textbf{Takeaway:} Attack performs better when starting with most selective attributes first. The low selectivity of \texttt{age} and \texttt{zip} introduces timing variance which results in loss of accuracy.}
\label{table:tuplext-reconstruction}
\begin{tabular}{rrrrr}
\toprule
\textbf{Ordering} & \multicolumn{1}{c}{\textbf{Queries} ($\times 10^6$)} & \multicolumn{1}{c}{\textbf{Recall}} & \multicolumn{1}{c}{\textbf{Precision}} & \multicolumn{1}{c}{\textbf{Time} (h)} \\
\midrule
\texttt{ssn}\,${\scriptstyle\to}$\,\texttt{zip}\,${\scriptstyle\to}$\,\texttt{age} & \textbf{\boldmath 14.3}\makebox[\widthof{\,\scalebox{0.7}{$\pm$\,\makebox[\widthof{13}][r]{13}.5}}][l]{\,\scalebox{0.7}{$\pm$\,\makebox[\widthof{13}][r]{}.0}} & \cellcolor[RGB]{0,104,55}\textcolor{white}{100.0\makebox[\widthof{\,\scalebox{0.7}{$\pm$\,\makebox[\widthof{17}][r]{17}.9}}][l]{\,\scalebox{0.7}{$\pm$\,\makebox[\widthof{17}][r]{}.0}}} & \cellcolor[RGB]{0,104,55}\textcolor{white}{100.0\makebox[\widthof{\,\scalebox{0.7}{$\pm$\,\makebox[0pt][r]{}.0}}][l]{\,\scalebox{0.7}{$\pm$\,\makebox[0pt][r]{}.0}}} & 1.3\makebox[\widthof{\,\scalebox{0.7}{$\pm$\,\makebox[\widthof{1}][r]{1}.5}}][l]{\,\scalebox{0.7}{$\pm$\,\makebox[\widthof{1}][r]{}.3}} \\
\texttt{ssn}\,${\scriptstyle\to}$\,\texttt{age}\,${\scriptstyle\to}$\,\texttt{zip} & \textbf{\boldmath 14.3}\makebox[\widthof{\,\scalebox{0.7}{$\pm$\,\makebox[\widthof{13}][r]{13}.5}}][l]{\,\scalebox{0.7}{$\pm$\,\makebox[\widthof{13}][r]{}.0}} & \cellcolor[RGB]{0,104,55}\textcolor{white}{100.0\makebox[\widthof{\,\scalebox{0.7}{$\pm$\,\makebox[\widthof{17}][r]{17}.9}}][l]{\,\scalebox{0.7}{$\pm$\,\makebox[\widthof{17}][r]{}.0}}} & \cellcolor[RGB]{0,104,55}\textcolor{white}{100.0\makebox[\widthof{\,\scalebox{0.7}{$\pm$\,\makebox[0pt][r]{}.0}}][l]{\,\scalebox{0.7}{$\pm$\,\makebox[0pt][r]{}.0}}} & 1.2\makebox[\widthof{\,\scalebox{0.7}{$\pm$\,\makebox[\widthof{1}][r]{1}.5}}][l]{\,\scalebox{0.7}{$\pm$\,\makebox[\widthof{1}][r]{}.1}} \\
\midrule
\texttt{zip}\,${\scriptstyle\to}$\,\texttt{ssn}\,${\scriptstyle\to}$\,\texttt{age} & 54.8\makebox[\widthof{\,\scalebox{0.7}{$\pm$\,\makebox[\widthof{13}][r]{13}.5}}][l]{\,\scalebox{0.7}{$\pm$\,\makebox[\widthof{13}][r]{}.0}} & \cellcolor[RGB]{0,104,55}\textcolor{white}{100.0\makebox[\widthof{\,\scalebox{0.7}{$\pm$\,\makebox[\widthof{17}][r]{17}.9}}][l]{\,\scalebox{0.7}{$\pm$\,\makebox[\widthof{17}][r]{}.0}}} & \cellcolor[RGB]{0,104,55}\textcolor{white}{100.0\makebox[\widthof{\,\scalebox{0.7}{$\pm$\,\makebox[0pt][r]{}.0}}][l]{\,\scalebox{0.7}{$\pm$\,\makebox[0pt][r]{}.0}}} & 5.0\makebox[\widthof{\,\scalebox{0.7}{$\pm$\,\makebox[\widthof{1}][r]{1}.5}}][l]{\,\scalebox{0.7}{$\pm$\,\makebox[\widthof{1}][r]{}.9}} \\
\texttt{zip}\,${\scriptstyle\to}$\,\texttt{age}\,${\scriptstyle\to}$\,\texttt{ssn} & 41.2\makebox[\widthof{\,\scalebox{0.7}{$\pm$\,\makebox[\widthof{13}][r]{13}.5}}][l]{\,\scalebox{0.7}{$\pm$\,\makebox[\widthof{13}][r]{13}.5}} & \cellcolor[RGB]{234,87,57}\textcolor{white}{45.6\makebox[\widthof{\,\scalebox{0.7}{$\pm$\,\makebox[\widthof{17}][r]{17}.9}}][l]{\,\scalebox{0.7}{$\pm$\,\makebox[\widthof{17}][r]{17}.9}}} & \cellcolor[RGB]{0,104,55}\textcolor{white}{100.0\makebox[\widthof{\,\scalebox{0.7}{$\pm$\,\makebox[0pt][r]{}.0}}][l]{\,\scalebox{0.7}{$\pm$\,\makebox[0pt][r]{}.0}}} & 4.5\makebox[\widthof{\,\scalebox{0.7}{$\pm$\,\makebox[\widthof{1}][r]{1}.5}}][l]{\,\scalebox{0.7}{$\pm$\,\makebox[\widthof{1}][r]{1}.5}} \\
\midrule
\texttt{age}\,${\scriptstyle\to}$\,\texttt{ssn}\,${\scriptstyle\to}$\,\texttt{zip} & 30.1\makebox[\widthof{\,\scalebox{0.7}{$\pm$\,\makebox[\widthof{13}][r]{13}.5}}][l]{\,\scalebox{0.7}{$\pm$\,\makebox[\widthof{13}][r]{}.1}} & \cellcolor[RGB]{0,104,55}\textcolor{white}{100.0\makebox[\widthof{\,\scalebox{0.7}{$\pm$\,\makebox[\widthof{17}][r]{17}.9}}][l]{\,\scalebox{0.7}{$\pm$\,\makebox[\widthof{17}][r]{}.0}}} & \cellcolor[RGB]{0,104,55}\textcolor{white}{100.0\makebox[\widthof{\,\scalebox{0.7}{$\pm$\,\makebox[0pt][r]{}.0}}][l]{\,\scalebox{0.7}{$\pm$\,\makebox[0pt][r]{}.0}}} & 2.8\makebox[\widthof{\,\scalebox{0.7}{$\pm$\,\makebox[\widthof{1}][r]{1}.5}}][l]{\,\scalebox{0.7}{$\pm$\,\makebox[\widthof{1}][r]{}.2}} \\
\texttt{age}\,${\scriptstyle\to}$\,\texttt{zip}\,${\scriptstyle\to}$\,\texttt{ssn} & 65.4\makebox[\widthof{\,\scalebox{0.7}{$\pm$\,\makebox[\widthof{13}][r]{13}.5}}][l]{\,\scalebox{0.7}{$\pm$\,\makebox[\widthof{13}][r]{7}.9}} & \cellcolor[RGB]{253,181,103}71.4\makebox[\widthof{\,\scalebox{0.7}{$\pm$\,\makebox[\widthof{17}][r]{17}.9}}][l]{\,\scalebox{0.7}{$\pm$\,\makebox[\widthof{17}][r]{10}.9}} & \cellcolor[RGB]{0,104,55}\textcolor{white}{100.0\makebox[\widthof{\,\scalebox{0.7}{$\pm$\,\makebox[0pt][r]{}.0}}][l]{\,\scalebox{0.7}{$\pm$\,\makebox[0pt][r]{}.0}}} & 7.7\makebox[\widthof{\,\scalebox{0.7}{$\pm$\,\makebox[\widthof{1}][r]{1}.5}}][l]{\,\scalebox{0.7}{$\pm$\,\makebox[\widthof{1}][r]{}.3}} \\
\bottomrule
\end{tabular}
\end{table}

\subsection{Potential Mitigations}

\vspace{-3.5pt}
\heading{Prior work.} The standard approach for handling timing side-channels is to change data-dependent execution paths into ones that use \textit{data-independent} or \textit{constant-time} evaluation (e.g., \cite{almeida_verifying_2016,molnar_program_2006}). Alternatively, a less robust approach is to carefully add noise or delays  \cite{hu_reducing_1992,zhang_predictive_2011}. %
Both likely come at significant performance costs for databases (which have variable workloads) and require  estimation of worst-case query  times. 

Dar et al. \cite{dar_rls_2023} propose a mitigation for \textit{unique keys}, or keys with at most cardinality~1, which aims to equalize the runtime of \textit{all} indexed point queries (regardless of whether the queried key exists) by
forcing RLS policy evaluation even on nonexistent queries.
This is done by forcing such queries to output a dummy record on index accesses and executing the policy filter on the dummy. As all queries result in policy evaluation, this eliminates the timing gap. 
However, the mitigation applies only to \textit{unique} keys: generalizing it to non-unique attributes would require 
slowing all queries to the worst case cardinality. Furthermore, even for unique keys, performance degrades when workloads contain a high fraction of nonexistent queries (which are slowed by design).

\heading{Using composite indexes and subqueries.}
The timing side-channel arises when PostgreSQL can  push down an
attacker-chosen, \texttt{LEAKPROOF} predicate $\qry$ into an index scan,
evaluate it over the full table $\tble$, and only then apply the RLS policy $\pred$ as a late-stage filter. While we can apply $\pred$ first, it means that we cannot leverage any indexes on the queried columns of $\qry$. 

Recall that the RLS policy from~\Cref{sec:evaluation} enforces that the \texttt{current\_user} can only see patients with the same \texttt{site\_id}.
Our approach is to bind the predicate into the same index access as the attacker's predicate so the scan is confined to the querying tenant's rows and $\attr = v$ is never evaluated against another tenant's data. This requires us to make two changes. 
First,  we change the index on each $\attr$ to  a \textit{composite index} of the form $(\texttt{site\_id}, \attr)$:
\begin{lstlisting}[language=SQL,numbers=none,
    aboveskip=4pt,
    belowskip=3pt]    
CREATE INDEX ... ON patients(site_id, atr)
\end{lstlisting}
This index is sorted lexicographically by $(\texttt{site\_id}, \attr)$. For queries that involve both $\texttt{site\_id}$ and $\attr$, this structure means that index scans on $\attr$ are scoped to a contiguous region of $\texttt{site\_id}$s and the equality check on $\texttt{site\_id}$ is discharged as part of the index access. 

Second, we require that the policy's $\texttt{site\_id}$  predicate is visible to the query optimizer so it knows it can use the index. The lookup in the \textbf{join} policy is wrapped in a function body that the planner cannot inspect, so we replace the function wrapper with the subquery in the policy schema directly:
\begin{lstlisting}[language=SQL,numbers=none,
    aboveskip=4pt,
    belowskip=3pt]
CREATE POLICY doctor_read ON patients FOR SELECT USING site_id = (SELECT site_id FROM doctors WHERE user_name = current_user);
\end{lstlisting}
As it depends only on the current user (a constant value), the optimizer can evaluate this subquery once per query that requires the index. We verified that this eliminates the structural side-channel by using  \texttt{EXPLAIN} to determine the query plan.

\begin{table}
    \caption{Our mitigation on the \Cref{table:oracle-num-queries} benchmark. \sethlcolor{yellow!0}\hl{\textbf{Takeaway:} The \emph{subquery $+$ composite index}  approach closes the side-channel for our scenario;  the oracle can only make random guesses. Adding just a composite index \emph{or} only changing the policy to a subquery is not sufficient to mitigate the~attack.}}
    \centering
    \begin{tikzpicture}[x=2.0cm,y=0.37cm]
\definecolor{rlsOuter}{RGB}{34,34,34}
\definecolor{rlsText}{RGB}{17,17,17}
\definecolor{rlsHeaderText}{RGB}{34,34,34}
\definecolor{rlsWhite}{RGB}{255,255,255}
\definecolor{rlsHighlight}{RGB}{255,250,205}
\useasboundingbox (0.000,1.000) rectangle (3.540,-4.000);
\node[font=\fontsize{6.5}{6.8}\selectfont, text=rlsHeaderText, inner sep=0pt, align=center] at (0.670,-0.000) {\textbf{Config}};
\node[font=\fontsize{6.5}{6.8}\selectfont, text=rlsHeaderText, inner sep=0pt, align=center] at (2.440,0.500) {\textbf{$k$ (Queries Per Probe Type)}};
\node[font=\fontsize{5.6}{6.0}\selectfont, text=rlsHeaderText, inner sep=0pt, align=center] at (1.615,-0.500) {1};
\node[font=\fontsize{5.6}{6.0}\selectfont, text=rlsHeaderText, inner sep=0pt, align=center] at (2.165,-0.500) {2};
\node[font=\fontsize{5.6}{6.0}\selectfont, text=rlsHeaderText, inner sep=0pt, align=center] at (2.715,-0.500) {4};
\node[font=\fontsize{5.6}{6.0}\selectfont, text=rlsHeaderText, inner sep=0pt, align=center] at (3.265,-0.500) {8};
\node[font=\fontsize{5.8}{6.2}\selectfont, text=rlsText, inner sep=0pt, align=center, anchor=west] at (0.060,-1.500) {join + composite};
\definecolor{rlsHeat0}{RGB}{165,0,38}
\path[fill=rlsHeat0, draw=none, line width=0pt] (1.340,-1.000) rectangle (1.890,-2.000);
\node[font=\fontsize{6.0}{6.5}\selectfont, text=rlsWhite, inner sep=0pt, align=center, anchor=east] at (1.640,-1.500) {100.0};
\node[font=\fontsize{4.3}{4.7}\selectfont, text=rlsWhite, inner sep=0pt, align=center, anchor=east] at (1.850,-1.500) {$\pm$\phantom{0}.0};
\definecolor{rlsHeat1}{RGB}{165,0,38}
\path[fill=rlsHeat1, draw=none, line width=0pt] (1.890,-1.000) rectangle (2.440,-2.000);
\node[font=\fontsize{6.0}{6.5}\selectfont, text=rlsWhite, inner sep=0pt, align=center, anchor=east] at (2.190,-1.500) {100.0};
\node[font=\fontsize{4.3}{4.7}\selectfont, text=rlsWhite, inner sep=0pt, align=center, anchor=east] at (2.400,-1.500) {$\pm$\phantom{0}.0};
\definecolor{rlsHeat2}{RGB}{165,0,38}
\path[fill=rlsHeat2, draw=none, line width=0pt] (2.440,-1.000) rectangle (2.990,-2.000);
\node[font=\fontsize{6.0}{6.5}\selectfont, text=rlsWhite, inner sep=0pt, align=center, anchor=east] at (2.740,-1.500) {100.0};
\node[font=\fontsize{4.3}{4.7}\selectfont, text=rlsWhite, inner sep=0pt, align=center, anchor=east] at (2.950,-1.500) {$\pm$\phantom{0}.0};
\definecolor{rlsHeat3}{RGB}{165,0,38}
\path[fill=rlsHeat3, draw=none, line width=0pt] (2.990,-1.000) rectangle (3.540,-2.000);
\node[font=\fontsize{6.0}{6.5}\selectfont, text=rlsWhite, inner sep=0pt, align=center, anchor=east] at (3.290,-1.500) {100.0};
\node[font=\fontsize{4.3}{4.7}\selectfont, text=rlsWhite, inner sep=0pt, align=center, anchor=east] at (3.500,-1.500) {$\pm$\phantom{0}.0};
\node[font=\fontsize{5.8}{6.2}\selectfont, text=rlsText, inner sep=0pt, align=center, anchor=west] at (0.060,-2.500) {subquery only};
\definecolor{rlsHeat4}{RGB}{165,0,38}
\path[fill=rlsHeat4, draw=none, line width=0pt] (1.340,-2.000) rectangle (1.890,-3.000);
\node[font=\fontsize{6.0}{6.5}\selectfont, text=rlsWhite, inner sep=0pt, align=center, anchor=east] at (1.640,-2.500) {100.0};
\node[font=\fontsize{4.3}{4.7}\selectfont, text=rlsWhite, inner sep=0pt, align=center, anchor=east] at (1.850,-2.500) {$\pm$\phantom{0}.0};
\definecolor{rlsHeat5}{RGB}{165,0,38}
\path[fill=rlsHeat5, draw=none, line width=0pt] (1.890,-2.000) rectangle (2.440,-3.000);
\node[font=\fontsize{6.0}{6.5}\selectfont, text=rlsWhite, inner sep=0pt, align=center, anchor=east] at (2.190,-2.500) {100.0};
\node[font=\fontsize{4.3}{4.7}\selectfont, text=rlsWhite, inner sep=0pt, align=center, anchor=east] at (2.400,-2.500) {$\pm$\phantom{0}.0};
\definecolor{rlsHeat6}{RGB}{165,0,38}
\path[fill=rlsHeat6, draw=none, line width=0pt] (2.440,-2.000) rectangle (2.990,-3.000);
\node[font=\fontsize{6.0}{6.5}\selectfont, text=rlsWhite, inner sep=0pt, align=center, anchor=east] at (2.740,-2.500) {100.0};
\node[font=\fontsize{4.3}{4.7}\selectfont, text=rlsWhite, inner sep=0pt, align=center, anchor=east] at (2.950,-2.500) {$\pm$\phantom{0}.0};
\definecolor{rlsHeat7}{RGB}{165,0,38}
\path[fill=rlsHeat7, draw=none, line width=0pt] (2.990,-2.000) rectangle (3.540,-3.000);
\node[font=\fontsize{6.0}{6.5}\selectfont, text=rlsWhite, inner sep=0pt, align=center, anchor=east] at (3.290,-2.500) {100.0};
\node[font=\fontsize{4.3}{4.7}\selectfont, text=rlsWhite, inner sep=0pt, align=center, anchor=east] at (3.500,-2.500) {$\pm$\phantom{0}.0};
\path[fill=rlsHighlight, draw=none, line width=0pt] (0.000,-3.000) rectangle (1.340,-4.000);
\node[font=\fontsize{5.8}{6.2}\selectfont, text=rlsText, inner sep=0pt, align=center, anchor=west] at (0.060,-3.500) {subquery + composite};
\definecolor{rlsHeat8}{RGB}{1,106,56}
\path[fill=rlsHeat8, draw=none, line width=0pt] (1.340,-3.000) rectangle (1.890,-4.000);
\node[font=\fontsize{6.0}{6.5}\selectfont, text=rlsWhite, inner sep=0pt, align=center, anchor=east] at (1.640,-3.500) {49.9};
\node[font=\fontsize{4.3}{4.7}\selectfont, text=rlsWhite, inner sep=0pt, align=center, anchor=east] at (1.850,-3.500) {$\pm$2.8};
\definecolor{rlsHeat9}{RGB}{3,110,58}
\path[fill=rlsHeat9, draw=none, line width=0pt] (1.890,-3.000) rectangle (2.440,-4.000);
\node[font=\fontsize{6.0}{6.5}\selectfont, text=rlsWhite, inner sep=0pt, align=center, anchor=east] at (2.190,-3.500) {51.5};
\node[font=\fontsize{4.3}{4.7}\selectfont, text=rlsWhite, inner sep=0pt, align=center, anchor=east] at (2.400,-3.500) {$\pm$2.8};
\definecolor{rlsHeat10}{RGB}{5,113,60}
\path[fill=rlsHeat10, draw=none, line width=0pt] (2.440,-3.000) rectangle (2.990,-4.000);
\node[font=\fontsize{6.0}{6.5}\selectfont, text=rlsWhite, inner sep=0pt, align=center, anchor=east] at (2.740,-3.500) {52.5};
\node[font=\fontsize{4.3}{4.7}\selectfont, text=rlsWhite, inner sep=0pt, align=center, anchor=east] at (2.950,-3.500) {$\pm$2.8};
\definecolor{rlsHeat11}{RGB}{0,104,55}
\path[fill=rlsHeat11, draw=none, line width=0pt] (2.990,-3.000) rectangle (3.540,-4.000);
\node[font=\fontsize{6.0}{6.5}\selectfont, text=rlsWhite, inner sep=0pt, align=center, anchor=east] at (3.290,-3.500) {49.0};
\node[font=\fontsize{4.3}{4.7}\selectfont, text=rlsWhite, inner sep=0pt, align=center, anchor=east] at (3.500,-3.500) {$\pm$2.8};
\draw[rlsOuter, line width=0.7pt] (0.000,1.000) -- (3.540,1.000);
\draw[rlsOuter, line width=0.4pt] (1.390,-0.000) -- (3.490,-0.000);
\draw[rlsOuter, line width=0.4pt] (0.000,-1.000) -- (3.540,-1.000);
\draw[rlsOuter, line width=0.7pt] (0.000,-4.000) -- (3.540,-4.000);
\end{tikzpicture}
    \label{fig:mitigation}
\end{table}

\heading{Microbenchmark.} We ran our attack benchmarks from \Cref{sec:oracle-strength} on our composite-index-based mitigation. The experimental setup is identical. In \Cref{fig:mitigation}, we see that with the composite index enabled, the timing differences are indistinguishable (as shown by the CI), as the query's use of composite indexes structurally forces the evaluation of \texttt{site\_id} before the queried $\attr$. We also see that both changes are necessary---only changing the index format or the policy (but not the other) still is vulnerable to the attack.

\heading{Limitations.}
Composite tenant-leading indexes (with the policy rewrite) eliminate the timing side-channel for this kind of policy, but they are not a general replacement for RLS. They do not address policies that cannot be expressed as leading-index constraints.
Policies depending on complex per-row logic 
may not allow for a simple policy.
Also, composite indexes  may be
impractical for tables with many indexed attributes. In such cases, operators must choose
which attributes warrant composite indexing or accept higher query costs. It also slows down queries for operators who are not subject to RLS, since $\attr$ is no longer contiguous.

\subsection{Discussion}

We showed that, with a feasible number of queries, attackers can bypass
RLS and recover entire rows with high fidelity. Our experiments show that the
complications arising from complex queries and network noise can be overcome.
In practice, it may not be easy to exploit this side-channel when system activity
obscures the timings or queries are rate-limited. For the latter issue, we
stress that a fairly small number of queries is required to extract a single
row, and this can be targeted (e.g., if the adversary is targeting a person with a specific SSN). We conclude that a careful evaluation of
this channel is critical to secure RLS-protected databases.

\section{\colorbox{magenta!10}{DLS Scoring-Based Attacks}}
\label{sec:dls}

\setlength{\abovedisplayskip}{4.5pt}
\setlength{\belowdisplayskip}{4.5pt} 
\setlength{\abovedisplayshortskip}{2pt} 
\setlength{\belowdisplayshortskip}{2pt}  

We now present our second case study: attacks on \textit{document-level security} (DLS)
mechanisms in full-text search systems, specifically Elasticsearch/OpenSearch
(ES/OS). (Since both systems are built on a common foundation \cite{lucene} and OS
is a fork of ES, they work essentially identically for our purposes.) We start
with the relevant background, describe how relevant query types work, and then
describe and evaluate our attack which exploits a new \textit{prefix-expansion} side-channel in ES/OS.

\subsection{Background and Prior Work}\label{sec:dls-background}

\heading{Notation.} A \emph{document} $\doc$ contains text \textit{fields}, each
of which contain  an (ordered) sequence of \textit{terms}. (For ease of
exposition, we assume each $\doc$ contains a single text field.)  A \emph{document corpus}
$\corp$ is a set of documents. The set of terms in a document is determined
by applying a \textit{text analyzer} algorithm $\Analyze$ at indexing time,
which normalizes the text (e.g., lowercase, stemming, stopword removal) and
segments raw text into indexable terms. Our attacks  target the
\emph{analyzed} view of a document, so depending on the analyzer, some
components of the original text may be unrecoverable---for example, the default
\texttt{lowercase} analyzer removes capitalization. We denote the space
character by $\spc$ and  concatenation  by $\Concat$.

\heading{Instantiating the abstract FGAC model.} ES/OS hold a document corpus
that is protected by post-filtering DLS that limits an adversary to only viewing
documents that it owns. The adversary can create new documents that will show up
as results in future queries.  The adversary can issue text search queries from
an expressive set supported by the software.  We assume the adversary knows the
set of characters that may appear in the data (e.g., ASCII characters). We do not
assume the adversary knows anything further about the distribution or
composition of the corpus.  For ease of exposition, we  assume that the adversary
knows  deployment-specific configurations (e.g., the indexes and schema) though, in all cases, an adversary can discover these
settings through simple probing.

The architecture and basic idea of the side-channel are described in
\Cref{fig:overview}. The response to a query includes a list of
documents along with relevance scores that can be used to rank them (below we
discuss how these scores are computed).  The list and scores are produced by a
system that  does not take access permissions into account. Instead, a filter
is configured with a predicate $\pred$ that encodes access privileges, so only
documents satisfying $\pred$ are returned.  The predicate $\pred$ may depend on
per-document fields and even other system information (for example, $\pred$ may
enforce that users can only access documents tagged with their name, or
documents containing a keyword in a  text field).

\heading{BM25 scoring.}  Ranking documents via  
\textit{scoring} is a standard feature of full-text search systems. 
We recall the relevant details of scoring in ES/OS, where
several query types use a variation of BM25 scoring~\cite{robertson_probabilistic_2009} as implemented in Lucene~\cite{lucene}. BM25  depends on configurable constants $k_1, b$, and $\boost$. %
We omit $\boost$ as it has no effect  when $\boost = 1$ (the default).

Given a document $\doc$, query $q = t_1\spc t_2\spc \ldots$ consisting of one or
more terms, and a corpus $\corp$, we denote the  score by
$\scorebm(\doc,q,\corp)$. This score depends on the \emph{term frequency (TF)} of
each term $t\in\{t_1,t_2,\ldots\}$ with respect to $\doc$, denoted $\tf(t,\doc)$, which is the
number of times $t$ appears in $\doc$. It also depends on the \emph{document
frequency (DF)} of $t$ with respect to the entire corpus $\corp$, or the
number of documents containing $t$:
\[ \df(t,\corp) = |\{\doc\in\corp : t\in\doc\}|.\] 
With these values, the score $\scorebm(\doc,q,\corp)$ is computed as
\begin{align}
  \scorebm(\doc,q,\corp) & = \sum_{t\in q} \idf(t,\corp)\cdot
    \mathrm{tfNorm}(t,\doc,\corp). \label{eq:score}
\end{align}
The  $\mathrm{tfNorm}$ values in this sum are computed as
\begin{align*}
\mathrm{tfNorm}(t,\doc,\corp) & = \frac{\tf(t,\doc)}
    {\tf(t,\doc)+k_1
      \Bigl(1 - b + b\cdot\frac{\dl(\doc)}{\avgdl(\corp)}\Bigr)}
\end{align*}
where $\dl(\doc)$ is the length of $\doc$, $\avgdl(\corp)$ is the average length
of documents in the corpus.  The other
values are the \emph{inverse document frequencies} $\idf(t,\corp)$, which are
computed via
\begin{equation}
  \idf(t,\corp)
  = \log\biggl(1 + \frac{N-\df(t,\corp) + 0.5}{\df(t,\corp) + 0.5} \biggr), \label{eq:idf}
\end{equation}
where $N$ is the number of documents in $\corp$.  
The intuition is that higher
$\tf$ values will boost the per-term scores, while higher $\df$ values will
lower them.

\heading{The DF side-channel.} In principle, a timing side-channel against
post-filtering DLS may exist, but we were unable to capture as reliable of a timing delta in preliminary experiments (compared to PostgreSQL). Thus, we use a completely different mechanism for attacking DLS: we exploit the fact that scores on attacker-visible documents \emph{depend on private documents}. In particular, given a term $t$, scores depend on the DF of $t$ which itself depends on corpus-wide 
documents.

This \textit{DF side-channel} is exploitable in
DLS systems where an attacker can issue  queries 
\textit{and} insert documents.  In this setting, B\"uttcher and Clarke~\cite{buttcher2005security} gave the
following attack.  If an attacker wants 
to test if a particular term $t$ is present in a private document, the attacker injects two documents: $\Ctrldoc$, which consists of a random term $\Ctrlr$ that is assumed to not appear in the index, and $\Canddoc$ containing the term of interest $t$. They then search for $\Ctrlr$, which returns $\Ctrldoc$ along with a
score $\Ctrls$.  Next, they search for $t$, which returns $\Canddoc$ with score $\Cands$.
If $\Ctrls \approx \Cands$,  they guess that $t$ is not present in any
private documents.  If $\Ctrls \gg \Cands$,  they guess that $t$ is 
present.  This  works because if $t$ is not present, then both $t$ and
$\Ctrlr$ will have DFs equal to $1$ as they are only present in the attacker-created 
 documents. Conversely, if $t$ is present in a private document, then its
higher DF results in a lower score $\Cands$.  

B\"uttcher and Clarke \cite{buttcher2005security} observed that this attack applied to
systems such as Apple Spotlight. Wang et al. \cite{wang_side-channel_2017} showed it was
practically relevant to several cloud-hosted systems, notably GitHub, 
and enumerated several challenges related to system noise and unknown system
configurations.

\heading{Prior limitations.} Much like the RLS attacks in \Cref{sec:rls-background}, the DF side-channel in \cite{buttcher2005security,wang_side-channel_2017}  only tests whether an exact indexed term exists. This requires the attacker to know candidate terms in advance and 
 cannot recover high-entropy
terms such as keys, passwords, or long, arbitrary strings in general.  Nevertheless, we will show these limits
disappear once we account for ES/OS's expressive query interfaces.

\subsubsection{ES/OS Features and Queries}\label{sec:dls-queries}

We now describe several query types in ES/OS that we exploit.

\heading{\match  queries.} Basic full-text search is supported via
\match queries. A $\match(q)$ query transforms the input string $q$ into an
unordered set of terms $\{t_i~|~t_i \in \Analyze(q)\}$, and outputs documents that contain at least one of $\{t_1, \dots, t_r\}$. BM25 scores are used to rank the results.

\newcommand{\CatExample}{\HLColor{purple!15}{cat}}
\newcommand{\DogExample}{\HLColor{green!15}{dog}}
\newcommand{\FExample}{\HLColor{brown!25}{f}}
\heading{\matchphraseprefix (\mpp) queries.} ES/OS support  autocomplete
functionality via \matchphraseprefix (abbreviated as \mpp) queries. A $\mpp(q)$
query  transforms $q$ into an \textit{ordered} sequence of terms $t_1\spc
\dots\spc t_r \gets \Analyze(q)$.  The first $r-1$ terms are treated as a 
\emph{phrase} $\phrase=t_1\spc\ldots\spc t_{r-1}$, and the final term $t_r$ 
 is treated as the \emph{prefix}; below we denote such a query by
 $q=(\phrase,t_r)$.  $\mpp$ only returns documents that
contain 
$\phrase$ \emph{in order} followed by a term starting with $t_r$.\footnote{For example,
$\mpp(\text{``\texttt{\CatExample\spc \DogExample\spc \FExample{}}''})$ will match documents containing  ``\texttt{\CatExample\spc \DogExample \spc \FExample{}}'', ``\texttt{\CatExample\spc \DogExample\spc \FExample{}ox}'', or ``\texttt{\CatExample\spc \DogExample\spc \FExample{}erret}'', but not a document containing
``\texttt{\DogExample \spc \CatExample\spc \FExample{}erret}'' or ``\texttt{\CatExample \spc \DogExample\spc rand\spc \FExample{}erret}''.}

BM25 scores for \mpp queries are  computed in one of two ways depending on whether or not the schema specifies that the text field has a \textit{prefix index}, which speeds up prefix queries in exchange for additional storage. (Normal text fields do not have a prefix index by default.) 

When there is no prefix index, scores are computed as follows. Given a query
$q=(\phrase,t_r)$ and a document $\doc$, let 
$\pf(q, \doc)$  be the \emph{phrase frequency} of $q$, or the number of appearances in $\doc$
of the complete, ordered set of terms in $q$ ($\phrase$ followed by  $t_r$) ; also, let $\exp(t_r,\corp)$
be the set of \emph{expansions} of $t_r$, which are indexed terms starting with
$t_r$ (including $t_r$ itself, if it is present).  The maximum number
of expansions in the sum is capped by a \texttt{max\_expansions} parameter (by
default, to 50). Specifically, the set $\expand(t_r,\corp)$ is the first
\texttt{max\_expansions} expansions in alphabetical order. 
The score
is computed as
\begin{align}\label{eq:idf-mpp-no-index}
  \pf(q,\doc) \cdot \bigg( \sum_{t\in \phrase} \idf(t,\corp) 
  + \sum_{t'\in \expand(t_r,\corp)} \idf(t',\corp)\bigg).
\end{align}
When there is a prefix index, the score is computed as in \Cref{eq:idf-mpp-no-index},
except that
the second $\idf$ summation is replaced with a single alternative value,
$\idfprefix(t_r,\corp)$.
$\idfprefix$ is computed as in \Cref{eq:idf}, except that $\df(\cdot)$
is replaced with
\[ \dfprefix(t_r,\corp) = 
|\{\doc\in\corp : \textrm{$\doc$ contains a term with prefix $t_r$} \}|.\] 
The score is now 
computed as
\begin{align}\label{eq:idf-mpp-index}
  \pf(q,\doc) \cdot \bigg( \sum_{t\in \phrase} \idf(t,\corp) 
  + \idfprefix(t_r,\corp)\bigg).
\end{align}

\heading{The prefix-expansion (PE) side-channel.} With or without a prefix
index, we observe that this type of scoring creates a new \emph{prefix-expansion
(PE) side-channel}. Consider when there is no prefix index: the second summation
in (\ref{eq:idf-mpp-no-index}) is over the set $\expand(t_r,\corp)$, which
depends on private documents.  The contributions of the terms are still
affected as in the DF side-channel, but we will crucially exploit the \emph{number
of terms in the sum} (for the first time). In particular, our attacker will detect
when this set is empty or not.  With prefix indexing,
the effect is similar, except curiously the effect on the scoring is
reversed---prefix existence  causes the score to decrease, not increase.  This
is because in (\ref{eq:idf-mpp-index}), $\idfprefix(t_r,\corp)$ decreases as the number of expansions
increases.  In either case, prefix information is not
present in the original DF side channel, and it is what enables our
stronger attack.

\subsection{Bag-of-Terms Extraction}\label{sec:bag-extraction}

We now show how to exploit the DF and PE side-channels to recover high-entropy, unknown index terms, which was not possible in prior work. To do this, we will construct two  
oracles that determine if (1) a chosen term $t$ appears in a given  $\mathsf{index}$ and (2) a term in $\mathsf{index}$ begins with a chosen prefix $p$. 

\begin{algorithm}[t]
\footnotesize
   \SetKwInOut{Input}{Input}
   \SetKwInOut{Output}{Output} 
   \SetKwInOut{Parameter}{Parameter} 

   \caption{\textsf{ExactOracle}}
   \label{alg:existence-oracle}
   \KwIn{Term $t$, index name $\mathsf{index}$}
    \KwOut{\textsf{True} iff $t$ is an indexed term in $\mathsf{index}$}
   $\Ctrlr \gets \rndterm$\;
   $\Write(\Canddoc, t)$\;
   $\Write(\Ctrldoc, \Ctrlr)$\;
   $(\_,\Cands) \gets \match(\mathsf{index},t)$\;
   $(\_,\Ctrls) \gets \match(\mathsf{index},\Ctrlr)$\;
   \Return{$\Ctrls > \Cands$}\;
\end{algorithm}

\heading{Membership oracle.} The first oracle is a generalization of the prior existence oracles~\cite{buttcher2005security,wang_side-channel_2017} and is given in
Algorithm~\ref{alg:existence-oracle}. It checks if $t$ appears as an indexed term in
the corpus $\corp$. It starts by sampling an \textit{exact-fresh} term $\Ctrlr$. By \textit{exact-fresh}, we mean that $\Ctrlr$ does not exist in $\corp$, which we can do by using characters outside of $\corp$\footnote{The characters for fresh terms must be characters that are stable under text analysis (e.g., not removed). For example, in our experiments (\Cref{sec:dls-evaluation}) against the  Enron corpus (an English dataset), we generate fresh terms from the set of Cyrillic characters. } and ensuring we never reuse a term. Then, it writes $\Ctrlr$ to $\Ctrldoc$ and  writes $t$ to $\Canddoc$. It
then issues two \match queries for $t$ and $\Ctrlr$, and retrieves their
respective documents' scores $\Cands$ and $\Ctrls$ (it ignores the actual
results). If $\Ctrls > \Cands$, it concludes that $t$ was in the index. This  works because $\df(\Ctrlr, \corp) = 1$ (by the freshness of $\Ctrlr$), and $\df(t, \corp) \geq 1$ (since it is in $\Canddoc$); if $t$ already exists in the index, then $\df(t, \corp) > 1$, and then $\Cands < \Ctrls$.

\begin{algorithm}[t]
\footnotesize
   \SetKwInOut{Input}{Input}
   \SetKwInOut{Output}{Output} 
   \SetKwInOut{Parameter}{Parameter} 

   \caption{\textsf{PrefixOracle}}
   \label{alg:prefixoracle}
   \KwIn{Prefix $p$, index name $\mathsf{index}$}
   \KwOut{\textsf{True} iff $p$ is a prefix of a term in $\mathsf{index}$}
   $\Ctxr \gets \rndterm$\;
   $\Ctrlr \gets \textsc{prefixfreshterm}(|p|)$ \tcp*{$|p| = |\Ctrlr|$}
   $\sigma \gets$ character outside attack alphabet\;
   $\Write(\Canddoc, \Ctxr\spc p \Concat \sigma)$\;
   $\Write(\Ctrldoc, \Ctxr\spc \Ctrlr)$\;
   $(\_,\Cands) \gets \mpp(\mathsf{index},\Ctxr\spc p)$\;
   $(\_,\Ctrls) \gets \mpp(\mathsf{index},\Ctxr\spc \Ctrlr)$\;
   \lIf{$\mpp$ is using a prefix index}
     {\Return{$\Cands < \Ctrls$}}
     \lElse{\Return{$\Cands > \Ctrls$}}
\end{algorithm}

\heading{Prefix oracle.} 
Algorithm~\ref{alg:prefixoracle} defines \textsf{PrefixOracle}. It consumes a prefix $p$  and determines if any term in $\corp$ begins with $p$. It inserts  two documents $\Canddoc$ and $\Ctrldoc$, where $\Canddoc$ contains an exact-fresh \textit{context} term $\Ctxr$ followed by $ p\Concat\sigma$, where $\sigma$ is a character that is outside the attack alphabet, and $\Ctrldoc$ contains the same context term $\Ctxr$ followed by a \textit{prefix-fresh}  \textit{control} term $\Ctrlr$. By \textit{prefix-fresh}, we mean that $\expand(\Ctrlr, \corp) = \varnothing$ before $\Ctrlr$ is injected. (Once again, we can do this by constructing $\Ctrlr$ from characters outside the corpus with slightly more complicated bookkeeping  to ensure that we are not reusing prefixes.)
Then, it issues \mpp queries for $\Ctxr\spc p$ and $\Ctxr\spc \Ctrlr$ and compares the scores $\Cands$ and $\Ctrls$. 
Then, if $\mpp$ is \textit{not} using a prefix index, it returns \textsf{True} if and only if $\Cands > \Ctrls$. If $\mpp$ \textit{is} using a prefix index, the direction of the conditional \textit{reverses}---it returns \textsf{True} if and only if $\Cands < \Ctrls$.

As a technical remark, even if the field has a prefix index on it, each prefix index in ES/OS has a \textit{maximum prefix length} $\mathrm{max}_\mathrm{prefix}$ (maximum is 20). Given a prefix query $p$ on a field that has a prefix index, if $|p| > \mathrm{max}_\mathrm{prefix}$, then  $\mpp$  cannot use the prefix index and it will default to the normal scoring path. If the field has a prefix index, we can easily handle this in the oracle if we know $\mathrm{max}_\mathrm{prefix}$---by requiring that $|p| = |\Ctrlr|$, both $\mpp$ queries use identical scoring paths and we just need to flip the direction of our score check when $|p|> \mathrm{max}_\mathrm{prefix}$.\footnote{If the adversary does not know $\mathrm{max}_\mathrm{prefix}$ \textit{a priori}, they can easily learn it by observing the behavior of scores on attacker-injected documents.}

\heading{Correctness (without prefix index).} Suppose $\mpp$ \textit{is not} using a prefix index. Since $p\Concat\sigma$ and $\Ctrlr$ are fresh terms,  $\df(p\Concat\sigma,\corp) = \df(\Ctrlr,\corp) = 1$. 
Then, consider the case when $p$ \emph{is not} a prefix of any existing  term in
$\corp$. After $\Canddoc$ and $\Ctrldoc$ are injected, then $\expand(p,\corp)
=\{p\Concat\sigma\}$ and $\expand(\Ctrlr,\corp) = \{\Ctrlr\}$. Then the
following hold:
\begin{align*}
  \df(p\Concat\sigma,\corp) & = \df(\Ctrlr,\corp), \\
\idf(p\Concat\sigma,\corp) & = \idf(\Ctrlr,\corp), \\
\idf(\Ctxr,\corp) + \idf(p\Concat\sigma,\corp) & = \idf(\Ctxr,\corp) + \idf(\Ctrlr,\corp), \\
\idf_{\mathrm{mpp}}(\Ctxr \spc p,\corp) & = \idf_{\mathrm{mpp}}(\Ctxr\spc \Ctrlr,\corp).
\end{align*}
As the rest of \Cref{eq:score} is constant, we have  $\Ctrls = \Cands$ and so it returns \textsf{False} in this case.

Now, suppose $p$ \emph{is} a prefix of some indexed term $p'$ (possibly $p =
p'$).  Then, $\expand(p,\corp) \supseteq \{p\Concat\sigma, p'\}$ and
$\expand(\Ctrlr,\corp) = \{\Ctrlr\}$. Then the following hold:
\begin{align*}
\df(p\Concat\sigma,\corp)  &= \df(\Ctrlr,\corp), \\
\idf(p\Concat\sigma,\corp) & = \idf(\Ctrlr,\corp), \\
\idf(\Ctxr,\corp) + \idf(p\Concat\sigma,\corp) & = \idf(\Ctxr,\corp)+ \idf(\Ctrlr,\corp), \\
\idf(\Ctxr,\corp) + \idf(p\Concat\sigma,\corp) & +\, \idf(p',\corp) \\ &> \idf(\Ctxr,\corp)+ \idf(\Ctrlr,\corp), \\
\idf_{\mathrm{mpp}}(\Ctxr\spc p,\corp)  & > \idf_{\mathrm{mpp}}(\Ctxr\spc \Ctrlr,\corp),
\end{align*}
These imply $\Cands > \Ctrls$, so it returns \textsf{True} in this case.

\heading{Correctness (with prefix index).} Suppose $\mpp$ \emph{is} using a
prefix index. By construction,  $\dfprefix(\Ctrlr,\corp) = 1$. Consider the case
when $p$ \emph{is not} a prefix of any existing  term in $\corp$. This means $\dfprefix(p,\corp) =1$, so $\dfprefix(p,\corp)  = \dfprefix(\Ctrlr,\corp)$. This implies $\idfprefix(p,\corp)  = \idfprefix(\Ctrlr,\corp)$, and as the rest of \Cref{eq:score} is constant, it follows that the $\mpp$ scores  $\Ctrls = \Cands$ and so it returns \textsf{False} in this case.
Now, suppose $p$ \emph{is} a prefix of some indexed term. Then, $\dfprefix(p,\corp) >1$ and $\dfprefix(p,\corp) > \dfprefix(\Ctrlr,\corp)$. This implies $\idfprefix(p,\corp) < \idfprefix(\Ctrlr,\corp)$, which itself means $\Cands < \Ctrls$, so it returns \textsf{True} in this case.

\heading{Complexity.} Both $\textsf{ExactOracle}$ and $\textsf{PrefixOracle}$
require injecting two documents and issuing two queries.
In theory, one can reuse a fixed control document $\Ctrldoc$ across probes and only 
rewrite the candidate document $\Canddoc$, reducing document writes by a constant factor. (We do not do this for ease of implementation and exposition.) We discuss further optimizations later in the section.

\begin{algorithm}[t]
\footnotesize
   \SetKwInOut{Input}{Input}
   \SetKwInOut{Output}{Output} 
   \SetKwInOut{Parameter}{Parameter} 

   \caption{$\textsf{EnumerateTerms}$}
   \label{alg:enumerate-terms}
   \KwIn{Set of allowed characters $\chars$, index name $\mathsf{index}$, initial queue prefixes $Q_\mathsf{init}$}
   \KwOut{$\What$, the indexed terms in $\mathsf{index}$}

    $\What \gets \{\}$\;
    Initialize queue $Q \gets \{ g \Concat c ~:~g \in Q_\mathsf{init}, c \in \chars \}$\;
    \While{$Q$ is not empty}{
        $p \gets Q.\pop()$\;
        \If{$\textsf{PrefixOracle}(\mathsf{index},p)$}
        {
            \lFor{$c\in \chars$}{$Q.\push(p\Concat c)$}
        }
        \lIf{$\textsf{ExactOracle}(\mathsf{index},p)$}
        {
            $\What\gets\What\cup\{p\}$
        }
   }
   \Return $\What$
\end{algorithm}

\heading{Combining the oracles.} By combining  $\textsf{ExactOracle}$ and $\textsf{PrefixOracle}$, we enumerate all of the  terms  in
Algorithm~\ref{alg:enumerate-terms}. It consumes a set of characters that are allowed to appear in terms $\chars$, the index to reconstruct, and an initial queue state (for now, assume that $Q_\mathsf{init} =\{\texttt{""}\}$ is the empty string---we leave this parameterized as we  use this oracle as a subroutine for  further extraction in \Cref{sec:richer-n-gram}).
At a high level, it performs a breadth-first search over  the trie of  terms starting from the empty string. It continues traversing if adding a character from $\chars$ produces another valid prefix, and uses the $\textsf{ExactOracle}$ to determine if prefixes are  complete terms. 

\heading{Correctness.}
$\textsf{EnumerateTerms}$ performs a trie traversal over the space of
candidate terms. Each oracle query either finds that a prefix cannot
extend to any hidden terms or confirms that the prefix
occurs. Thus, the algorithm exactly enumerates 
the set of hidden terms.

\heading{Complexity.} Let
$V$ denote the number of trie nodes (candidate prefixes) of the set of terms.
The enumeration performs $O(V)$ oracle calls (one membership test per visited
node, plus a prefix test to decide whether to descend), so the total cost is linear
in the size of the explored portion of the prefix trie.
Since each oracle call issues two scored queries and writes two attacker-controlled
documents and each trie node has two oracle calls ($\textsf{PrefixOracle}$ and $\textsf{ExactOracle}$), enumeration uses $4V$ scored queries and $4V$ document injections.

\subsection{$n$-Gram Extraction}\label{sec:richer-n-gram}

Our attack so far extracts the unordered set of terms from any indexed text
field.  In this section we assume the system uses some commonly-enabled special
indexes that are designed to further optimize autocomplete workloads (like \mpp
queries). By abusing these indexes, we can extract richer structure that allows
us to recover approximate document text.

\heading{Search-as-you-type.} ES/OS provide a specialized  
 index type called \texttt{search\_as\_you\_type} (SAYT) designed to support interactive
``search-as-you-type'' workloads.
This field type is widely recommended for applications such as email search and enterprise knowledge bases, where users expect partial
queries to return relevant results as they type.

A \texttt{search\_as\_you\_type} index augments a standard full-text field
with additional inverted indexes over contiguous $n$-grams. Specifically, given a SAYT index on the field \texttt{field}, 
for a term sequence $(t_1,t_2,\dots,t_m) \gets \Analyze(\cdot)$, the system indexes all
contiguous $n$-grams $(t_i,\dots,t_{i+n-1})$  in the subfield \texttt{field.\_$n$gram} (for fixed values of $n$).
ES/OS support SAYT indexes with $n$-grams up to a configurable $n\in \{2,3,4\}$.
Each $n$-gram is treated as an independent term with its own term frequency
and document frequency statistics. For example, the  $3$-grams of the document ``$\texttt{dog\spc
cat\spc cow\spc fox}$''  are (1) ``$\texttt{dog\spc cat\spc cow}$'' and (2) 
``$\texttt{cat\spc cow\spc fox}$''. In the $\texttt{\_3gram}$ index, these
$3$-grams will be treated as independent terms. Any given $3$-gram
$g$ will have a DF with respect to the corpus, and a TF with respect to an individual
document.

These $n$-gram indexes are used transparently by higher-level query types---most 
notably, prefix-expanding queries such as \texttt{match\_phrase\_prefix}.
From the perspective of scoring, $n$-grams behave identically to ordinary
terms: their document frequencies contribute to BM25-style relevance scores,
and are computed over the entire corpus.

\heading{Extracting $n$-grams.} As the $\ngram$ indexes  can be queried
independently like normal fields, the attack insights from
\Cref{sec:bag-extraction} still apply to those fields---just over $n$-grams
instead of individual terms. For example, when issuing an $\mpp$ query on
$\ngram$ with input $q=t_1\spc \ldots\spc t_r$, the IDF  is computed by first listing
all $n$-grams via a sliding window of $n$ terms across the query. All $n$-grams
contribute to the IDF $\idfmpp$ as in the first sum in
\Cref{eq:idf-mpp-no-index}. The final $n$-gram is also treated as a prefix of
$n$-grams, and expanded similarly to the second sum in
\Cref{eq:idf-mpp-no-index}. Finally, the score is computed similarly to basic
\mpp using these values.

\begin{algorithm}[t]
\footnotesize
   \SetKwInOut{Input}{Input}
   \SetKwInOut{Output}{Output} 
   \SetKwInOut{Parameter}{Parameter} 

   \caption{\textsf{$n$GramExactOracle}}
   \label{alg:ngramexactoracle}
   \KwIn{$n$-gram $t$, SAYT index name $\mathsf{index}$}
    \KwOut{\textsf{True} iff $t$ is an indexed $n$-gram in $\mathsf{index}$}
   \hl{$\Ctrlr^1,\cdots,\Ctrlr^{n} \gets \rndterm$}\;
   \hl{$\Ctrlr \gets \Ctrlr^1\spc \cdots\spc \Ctrlr^{n}$}\;
   \hl{$\mathsf{index} \gets \mathsf{index}.$\texttt{\_$n$gram}}\;
   $\dots$ \tcp*{Continues from Line 2 of Algorithm~\ref{alg:existence-oracle}}
\end{algorithm}

\begin{algorithm}[t]
\footnotesize
   \SetKwInOut{Input}{Input}
   \SetKwInOut{Output}{Output} 
   \SetKwInOut{Parameter}{Parameter} 

   \caption{\textsf{$n$GramPrefixOracle}}
   \label{alg:ngramprefixoracle}
   \KwIn{$n$-gram prefix $p$, SAYT index name $\mathsf{index}$}
   \KwOut{\textsf{True} iff $p$ is a prefix of an $n$-gram in $\mathsf{index}$}
   \hl{$\Ctxr^1,\cdots,\Ctxr^{n-1} \gets \rndterm$}\;
   \hl{$\Ctxr \gets \Ctxr^1\spc \cdots\spc \Ctxr^{n-1}$}\;
   \hl{$\mathsf{index} \gets \mathsf{index}.$\texttt{\_$n$gram}}\;
   $\dots$ \tcp*{Continues from Line 2 of Algorithm~\ref{alg:prefixoracle}}
\end{algorithm}

Because of this, we can extract $n$-grams by applying our attack from Algorithm~\ref{alg:enumerate-terms} directly on the SAYT $\texttt{field.\_ngram}$ indexes with a small modification to our oracles. Instead of sampling a single fresh term, we must construct \textit{fresh} $(n-1)$-grams, as illustrated in Algorithm~\ref{alg:ngramexactoracle} and Algorithm~\ref{alg:ngramprefixoracle}. Then, we can directly query against the $\ngram$ SAYT indexes instead of the main $\mathsf{index}$.

However, SAYT imposes one additional challenge---while $\mpp$ queries work as normal against the $\ngram$ SAYT indexes, our $\textsf{PrefixOracle}$ does not work out of the box when trying to recover unigrams. 
Recall that a normal text field with a prefix index builds the index on prefixes of \textit{single} terms, which is what allows us to isolate the scoring contributions from a given prefix $p$ in Algorithm~\ref{alg:prefixoracle} (provided we use the correct score direction), as
\begin{align*}
\score(\doc,q,\corp) =~&  \idf(\Ctxr,\corp) \cdot
\mathrm{tfNorm}(\Ctxr, \doc,\corp)~+ \\
& \idfprefix(p,\corp)
\cdot
\mathrm{tfNorm}(p, \doc, \corp).
\end{align*}
Conversely, SAYT builds a prefix index over $n$-grams, so the indexed prefix includes the preceding terms, resulting in
\begin{align*}
\score(\doc,q,\corp) = \idfprefix(\Ctxr\spc p,\corp)
\cdot
\mathrm{tfNorm}(\Ctxr\spc p,\doc, \corp).
\end{align*}
Note the contrast---in the normal-field case the prefix  $p$ is scored as its own summand, so its IDF is isolable, but under SAYT the prefix index fuses $\Ctxr$ with $p$ into a single $n$-gram, so $p$'s contribution cannot be separated.
Since $\Ctxr$ is random and private documents do not contain it, after injection $\dfprefix(\Ctxr\spc p,\corp) = 1$, even if hidden documents contain terms beginning with $p$. By construction, $\dfprefix(\Ctxr\spc \Ctrlr,\corp)=1$, so $\Cands = \Ctrls$ always. Furthermore, $\mpp$ requires two terms to perform any prefix expansion, so we cannot simply omit $\Ctxr$. This breaks the original oracle for extracting 1-grams.

Nevertheless, there are a few ways of solving this problem; we give one such formulation here. We can write a semantically equivalent query to normal  $\mpp$ queries that allows us to control the scoring calculation using \texttt{span} queries. Whereas $\mpp$ has flexibility as to how it calculates the score, a \texttt{span} query is a low-level positional query that lets us explicitly control the expansion behavior. 
Our modified algorithm is in Algorithm~\ref{alg:1gramexactoracle} and works identically to Algorithm~\ref{alg:prefixoracle} but, instead of issuing $\mpp$ queries, we issue the query
\[
\texttt{span\_near}(
    \texttt{span\_term}(\Ctxr), \texttt{span\_multi}(\texttt{prefix}(p)) )
\]
The $\texttt{span\_term}(\Ctxr)$ clause requires the exact term $\Ctxr$; the $\texttt{span\_multi}(\texttt{prefix}(p))$ clause requires a term that is prefixed by $p$; the $\texttt{span\_near}$ parameters specify the order in which these conditions must be satisfied. This is the same semantics as $\mpp$, but works around the SAYT scoring issue as the scoring function is not delegated to the SAYT indexes. With this formulation, the score is computed as described in \Cref{sec:bag-extraction}, which allows our oracle to enumerate the unigrams in the SAYT field.

\begin{algorithm}[t]
\footnotesize
   \SetKwInOut{Input}{Input}
   \SetKwInOut{Output}{Output} 
   \SetKwInOut{Parameter}{Parameter} 
    \SetAlgoLined
    \setcounter{AlgoLine}{4} %

   \caption{\textsf{1GramPrefixOracle}}
   \label{alg:1gramexactoracle}
   \KwIn{Prefix $t$, SAYT index name $\mathsf{index}$}
    \KwOut{\textsf{True} iff $t$ is an indexed $1$-gram in $\mathsf{index}$}
   $\dots$ \tcp*{Continues after Line 5 of Algorithm~\ref{alg:prefixoracle}}
   \hl{$(\_,\Cands) \gets \texttt{span\_near}(
    \texttt{span\_term}(\Ctxr), \texttt{span\_multi}(\texttt{prefix}(p)))$}\;
   \hl{$(\_,\Ctrls) \gets \texttt{span\_near}(
    \texttt{span\_term}(\Ctxr), \texttt{span\_multi}(\texttt{prefix}(\Ctrlr)))$}\;
   $\dots$ \tcp*{Continues from Line 8 of Algorithm~\ref{alg:prefixoracle}}
\end{algorithm}

With this, we define Algorithm~\ref{alg:enumerate-ngrams}, which works identically to Algorithm~\ref{alg:enumerate-terms} but we replace $\textsf{PrefixOracle}$ with $\textsf{$n$GramPrefixOracle}$ (and $\textsf{1GramPrefixOracle}$) and $\textsf{ExactOracle}$ with $\textsf{$n$GramExactOracle}$. The correctness and complexity arguments follow directly from the corresponding arguments for Algorithm~\ref{alg:enumerate-terms} so we omit them here.

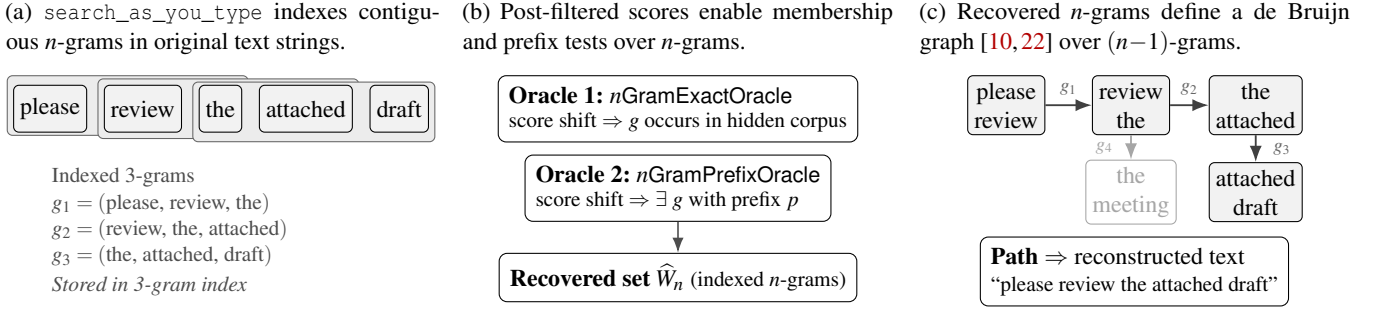
\begin{figure*}[t]
\centering
\setlength{\tabcolsep}{0pt}

\newcommand{\TokFill}{gray!10}
\newcommand{\AccentFill}{gray!15}
\newcommand{\Weak}{black!35}

\begin{subfigure}[t]{0.32\textwidth}
\centering
\caption{\texttt{search\_as\_you\_type} indexes contiguous $n$-grams in original text strings.}
\vspace{-2pt}
\begin{tikzpicture}[
  font=\small,
  baseline=(current bounding box.north),
  token/.style={draw, rounded corners=2pt, inner sep=2.5pt, minimum height=6.2mm, fill=\TokFill},
  gramhl/.style={fill=\AccentFill, draw=black!60, rounded corners=2pt},
  paneltitle/.style={font=\footnotesize, text=black!75},
  note/.style={font=\footnotesize, text=black!70},
  tight/.style={inner sep=0pt, outer sep=0pt},
]
\node[token] (t1) at (0,0) {please};
\node[token] (t2) [right=2.2mm of t1] {review};
\node[token] (t3) [right=2.2mm of t2] {the};
\node[token] (t4) [right=2.2mm of t3] {attached};
\node[token] (t5) [right=2.2mm of t4] {draft};

\begin{scope}[on background layer]
  \node[gramhl, fit=(t1)(t3), inner sep=2.2pt, yshift=1.2pt] {};
  \node[gramhl, fit=(t2)(t4), inner sep=2.2pt] {};
  \node[gramhl, fit=(t3)(t5), inner sep=2.2pt, yshift=-1.2pt] {};
\end{scope}

\node[note, anchor=west] (l0) at (-0.1,-0.9) {Indexed 3-grams};
\node[note, anchor=west] at (-0.1,-1.25) {$g_1=(\text{please, review, the})$};
\node[note, anchor=west] at (-0.1,-1.60) {$g_2=(\text{review, the, attached})$};
\node[note, anchor=west] at (-0.1,-1.95) {$g_3=(\text{the, attached, draft})$};

\node[note, anchor=west] at (-0.1,-2.35)
{\textit{Stored in  3-gram index}};
\end{tikzpicture}
\end{subfigure}
\hfill
\begin{subfigure}[t]{0.32\textwidth}
\centering
\caption{Post-filtered scores enable membership and prefix tests over $n$-grams.}
\vspace{-2pt}
\begin{tikzpicture}[
  font=\small,
  baseline=(current bounding box.north),
  card/.style={draw, rounded corners=3pt, inner sep=4.0pt, align=left, fill=white},
  pill/.style={draw, rounded corners=6pt, inner sep=2.0pt, fill=\TokFill},
  arr/.style={-Latex, line width=0.6pt, draw=black!70},
  tiny/.style={font=\scriptsize, text=black!65},
  lab/.style={font=\scriptsize, text=black!60},
]

\node[card, anchor=north] (o1) at (0,1.05)
{\textbf{Oracle 1:} \textsf{$n$GramExactOracle}\\[-2pt]
\footnotesize score shift $\Rightarrow$ $g$ occurs in hidden corpus};

\node[card, anchor=north] (o2) at (0,0)
{\textbf{Oracle 2:} \textsf{$n$GramPrefixOracle}\\[-2pt]
\footnotesize score shift $\Rightarrow$ $\exists~g$ with prefix $p$};

\node[card, anchor=north] (out) at (0,-1.3)
{\textbf{Recovered set} $\widehat{W}_n$ \footnotesize (indexed $n$-grams)};

\draw[arr] (o2.south) -- (out.north);

\end{tikzpicture}
\end{subfigure}
\hfill
\begin{subfigure}[t]{0.32\textwidth}
\centering
\caption{Recovered $n$-grams define a de Bruijn graph~\cite{bruijn_combinatorial_1946,good_normal_1946} over $(n\!-\!1)$-grams.}
\vspace{-2pt}
\begin{tikzpicture}[
  font=\small,
  baseline=(current bounding box.north),
  paneltitle/.style={font=\footnotesize, text=black!75},
  node/.style={draw, rounded corners=2pt, inner sep=2.5pt, fill=\TokFill, align=center},
  nodeweak/.style={draw=\Weak, text=\Weak, rounded corners=2pt, inner sep=2.5pt, fill=white, align=center},
  edge/.style={-Latex, line width=0.7pt, draw=black!75},
  edgeweak/.style={-Latex, line width=0.7pt, draw=\Weak},
  elab/.style={font=\scriptsize, text=black!60, fill=white, fill opacity=0.9, text opacity=1,
               inner sep=1.2pt, rounded corners=1pt},
  elabweak/.style={font=\scriptsize, text=\Weak, fill=white, fill opacity=0.9, text opacity=1,
                   inner sep=1.2pt, rounded corners=1pt},
  note/.style={font=\footnotesize, text=black!70},
  card/.style={draw, rounded corners=3pt, inner sep=4.0pt, align=left, fill=white},
]

\node[node] (a) at (-1.75,0.45) {please\\review};
\node[node] (b) at (-0.1,0.45) {review\\the};
\node[node] (c) at (1.55,0.45) {the\\attached};
\node[node] (d) at (1.55,-0.7) {attached\\draft};

\draw[edge] (a) -- (b) node[elab, midway, yshift=6pt] {$g_1$};
\draw[edge] (b) -- (c) node[elab, midway, yshift=6pt] {$g_2$};
\draw[edge] (c) -- (d) node[elab, midway, xshift=10pt] {$g_3$};

\node[nodeweak] (x) at (-0.1,-0.7) {the\\meeting};
\draw[edgeweak] (b) -- (x) node[elabweak, midway, xshift=-10pt] {$g_4$};

\node[card, align=left,anchor=west] at (-2.1,-1.75)
{\textbf{Path} $\Rightarrow$ reconstructed text\\[-1pt]
\footnotesize ``please review the attached draft''};

\end{tikzpicture}
\end{subfigure}

\caption{Overview of extracting reconstructable structure from DLS-protected \texttt{search\_as\_you\_type} indexes.}
\label{fig:dls-figure3}
\end{figure*}

\begin{algorithm}[t]
\footnotesize
   \SetKwInOut{Input}{Input}
   \SetKwInOut{Output}{Output} 
   \SetKwInOut{Parameter}{Parameter} 

   \caption{$\textsf{EnumerateNGrams}$}
   \label{alg:enumerate-ngrams}
    \SetAlgoLined
    \setcounter{AlgoLine}{3} %
   \KwIn{Set of allowed characters $\chars$, SAYT index name $\mathsf{index}$, integer $n$, initial queue prefixes $Q_\mathsf{init}$}
   \KwOut{$\What$, the indexed $n$-grams in $\mathsf{index}$}
  $\dots$ \tcp*{Continues after Line 4 of Algorithm~\ref{alg:enumerate-terms}}
        \If{\hl{$\textsf{$n$GramPrefixOracle}(\mathsf{index},p)$}} 
        {
            \lFor{$c\in \chars$}{$Q.\push(p\Concat c)$}
        }
        \lIf{\hl{$\textsf{$n$GramExactOracle}(\mathsf{index},p)$}}
        {
            $\What\gets\What\cup\{p\}$
        }
        $\dots$ \tcp*{Continues from Line 8 of Algorithm~\ref{alg:enumerate-terms}}
\end{algorithm}

\begin{algorithm}[t]
\footnotesize
   \SetKwInOut{Input}{Input}
   \SetKwInOut{Output}{Output} 
   \SetKwInOut{Parameter}{Parameter} 

   \caption{$\textsf{EnumerateAllGrams}$}
   \label{alg:enumerate-all-grams}
    \KwIn{Set of allowed characters $\chars$, SAYT index name $\mathsf{index}$, the length of $n$-grams to enumerate $n$}
    \KwOut{$\ngramshat$, the $n$-grams in $\corp$}
    $\What_1 \gets \textsf{EnumerateNGrams}(\chars,  \mathsf{index}, 1, \{\texttt{""}\})$\;
    \For{$j \in \{2, \dots, n\}$}{
        $Q_\mathsf{init}^j \gets \{g \Concat \spc \ : \ g\in \What_{j-1}\}$\;
        $\What_j \gets \textsf{EnumerateNGrams}(\chars, \mathsf{index}, j, Q_\mathsf{init}^j)$\;
    }
    \Return $\big(\What_1,\dots,\What_n\big )$\;

\end{algorithm}

\subsection{The Attack}

We now compose our scoring oracles into an end-to-end
reconstruction attack against DLS-protected document text.
At a high level, the attack proceeds in two stages:
(i) enumerating the set of $n$-grams present in the hidden corpus, and
(ii) reconstructing approximate document text from the recovered
$n$-grams. An overview of our attack is in \Cref{fig:dls-figure3}.

\heading{Enumerating $n$-grams.} We enumerate all of the $n$-grams in a
corpus using
Algorithm~\ref{alg:enumerate-all-grams}, which itself is composed of $\textsf{EnumerateNGrams}$. It uses $\textsf{EnumerateNGrams}$ to recover the set of all $1$-grams, then uses the set of recovered $1$-grams $\What_1$ to seed the starting search state of the next call to $\textsf{EnumerateNGrams}$ to recover the set of $2$-grams. In other words, it uses the discovered $(j-1)$-grams at each stage as prefixes to search for $j$-grams. This repeats until all $j$-grams have been discovered for $1 \leq j \leq n$.

Correctness follows directly from the corresponding argument for Algorithm~\ref{alg:enumerate-terms}; we omit a full analysis for brevity. For complexity, let 
$V_n$ be the number of candidate prefixes of the set of $n$-grams. The enumeration performs $O(V_n)$ oracle calls (linear
in the size of the  prefix trie) and requires $4V_n$ scored queries and $4V_n$ document injections. Once again, this follows directly from the complexity analysis for Algorithm~\ref{alg:enumerate-terms}.

\heading{Towards approximate document reconstruction.} 
Once we recover the set of indexed $n$-grams, we
can construct a
  \textit{de~Bruijn graph}~\cite{bruijn_combinatorial_1946,good_normal_1946} over $(n-1)$-grams: each $n$-gram $(t_1,\dots,t_n)$ defines a directed edge from the
$(n-1)$-gram prefix $(t_1,\dots,t_{n-1})$ to the suffix
$(t_2,\dots,t_n)$. Paths in this graph correspond to sequences of overlapping
$n$-grams and thus to candidate document substrings. 
When the graph is unambiguous, paths directly recover contiguous
sections of the original plaintext. At branch points, multiple
  reconstructions are consistent with the same $n$-gram set; we opt for a 
  greedy approach which identifies 
  longer, plausible (though possibly incorrect) reconstructions~\cite{pevzner_2001,galle_reconstructing_2015,hessel_how_2021}. We expand on this in \Cref{sec:dls-evaluation}.

\subsubsection{Additional Optimizations}\label{sec:dls-optimizations}

The algorithms above describe the attack in its simplest form, where each trie
node is tested independently. Here, we identify several  optimizations that reduce the adversarial overhead without changing the logical oracle being evaluated.

\heading{Packing and batching.} For a set of candidate prefixes
$p_1,\ldots,p_m$, instead of writing one candidate document and one control
document per prefix, we write one packed candidate document containing all
candidate probe phrases and one packed control document containing the matched
control phrases.\footnote{For
the SAYT $n$-gram enumeration, we additionally insert exact-fresh terms between
probe phrases so that the packed document does not accidentally create
additional $n$-grams spanning two adjacent probes. } As long as we  stay within ES/OS's maximum document size, one injection can prepare many oracle evaluations. This substantially reduces the number of injected documents. 

We further reduce the number of network round-trips with ES/OS's bulk insert and multi-search APIs. This does not change the number
of logical queries: each oracle call still has a candidate and control query. It only reduces the number of requests 
needed to execute those logical queries.

\heading{Leaf inference.} In term enumeration, we can reduce the number of $\textsf{ExactOracle}$ (and $\textsf{$n$GramExactOracle}$) calls by learning from previous $\textsf{PrefixOracle}(\cdot, p)$ calls. If
$\textsf{PrefixOracle}(\cdot, p) = \textsf{False}$, then no indexed term can equal $p$, so we can skip $\textsf{ExactOracle}(\cdot,p)$. If $\textsf{PrefixOracle}(\cdot, p) = \textsf{True}$, we can \textit{defer} the exact check on $p$ and instead check the  children of $p$ in the trie. If all children result in negative $\textsf{PrefixOracle}$ calls, then we can conclude $p$ is a leaf without an $\textsf{ExactOracle}$ call.
In other words, exact checks are only needed for internal trie
nodes; that is, terms that are also prefixes of longer terms.

\heading{$n$-gram trie.} For $n$-gram recovery we use the already recovered lower-order grams to
restrict the search space. After recovering unigrams, the
$2$-grams search only explores prefixes of recovered unigrams as possible
second terms. Then, the $j$-gram stage uses recovered $(j-1)$-grams as
contexts and the recovered unigram trie as the search space. This
avoids traversing
strings that cannot be valid next terms.

\begin{table}[t] 
\centering
\footnotesize
\caption{Attack cost separated by $n$-gram stages. The \textit{\# of injected documents} is $2\times$ query batches. The \textit{\# of trie nodes} is $0.5\times$ logical queries. Times are averaged over 3 repetitions. \textbf{Takeaway:} The  attacks recover the indexed $n$-gram set exactly across all tested corpus sizes. Injections and queries scale linearly with the \# of trie nodes.}
\begin{tabular}{rrrrrr}
\toprule
\bf \makecell[r]{\#\\Docs} & \bf Stage & \bf \makecell[r]{Indexed\\Terms} & \bf \makecell[r]{Logical\\ Queries} & \bf \makecell[r]{Inject/Query\\ Batches} & \bf \makecell[r]{Time\\(min)} \\
\midrule
\multirow{5}{*}{1} & 1-gram & 21 & 24{,}658 & 27 & 0.1 \\
 & 2-gram & 21 & 804 & 95 & $< 0.1$ \\
 & 3-gram & 20 & 798 & 94 & $< 0.1$ \\
 & 4-gram & 19 & 764 & 92 & $< 0.1$ \\
 & \textit{Total} &  & 27{,}024 & 308 & 0.2 \\
\addlinespace
\multirow{5}{*}{10} & 1-gram & 410 & 390{,}910 & 114 & 1.2 \\
 & 2-gram & 659 & 42{,}802 & 2{,}371 & 0.6 \\
 & 3-gram & 708 & 61{,}706 & 3{,}568 & 1.0 \\
 & 4-gram & 721 & 65{,}466 & 3{,}733 & 1.1 \\
 & \textit{Total} &  & 560{,}884 & 9{,}786 & 3.9 \\
\addlinespace
\multirow{5}{*}{100} & 1-gram & 2{,}875 & 2{,}587{,}220 & 660 & 9.5 \\
 & 2-gram & 9{,}506 & 580{,}564 & 22{,}428 & 7.0 \\
 & 3-gram & 12{,}204 & 1{,}276{,}596 & 63{,}432 & 18.8 \\
 & 4-gram & 13{,}098 & 1{,}531{,}738 & 79{,}596 & 24.3 \\
 & \textit{Total} &  & 5{,}976{,}118 & 166{,}116 & 59.6 \\
\addlinespace
\multirow{5}{*}{1000} & 1-gram & 10{,}179 & 8{,}972{,}494 & 2{,}293 & 39.4 \\
 & 2-gram & 46{,}843 & 3{,}355{,}842 & 91{,}987 & 36.8 \\
 & 3-gram & 68{,}364 & 8{,}508{,}622 & 355{,}953 & 114.3 \\
 & 4-gram & 76{,}809 & 10{,}983{,}702 & 498{,}256 & 170.5 \\
 & \textit{Total} &  & 31{,}820{,}660 & 948{,}489 & 361.0 \\[3pt]
\rowcolor[RGB]{0,104,55} \multicolumn{6}{>{\centering\arraybackslash}p{0.95\columnwidth}}{\textcolor{white}{\bf Across all corpus sizes and runs, each stage recovers the indexed $n$-gram set exactly (100\% recall and precision).}} \\
\bottomrule
\end{tabular}

\label{tab:dls-recovery}
\end{table}

\subsection{End-to-End Evaluation}\label{sec:dls-evaluation}

We now  evaluate the practicality of our scoring-based attacks
against {ES/OS}. Our evaluation aims to validate
the reliability of the scoring-based oracles we construct, analyze how
efficiently these oracles can be amplified to recover n-grams, and to what
extent 
recovered n-grams suffice to reconstruct documents. We ran our
attacks on OS 3.6.0 and ES 9.3.3.

\heading{Setup.}  We deploy ES/OS with DLS enabled and default BM25 scoring as a single-shard index to avoid confounding effects from cross-shard score normalization, though we remark there are known techniques to handle this \cite{wang_side-channel_2017}. We run the DBMS on a  \texttt{c4-standard-16} GCE host with  16 vCPUs (dedicated cores), 60 GB RAM, and a 200 GB local SSD. The attacker is running on a \texttt{c4-standard-16} host in the same zone and connects to the database as a DBMS user that can issue DLS-protected read queries on the collection. Our attack is implemented as a single-threaded program in Python 3.10 and includes the optimizations from \Cref{sec:dls-optimizations}. We explicitly refresh the index after each injection, though the attack still works without refreshes if the adversary can wait~\cite{wang_side-channel_2017}. 

\heading{Schema and data.} The corpus consists of a variable number of emails drawn from the Enron dataset \cite{klimt_introducing_2004}, a widely used, real-world dataset containing naturally occurring English text with heterogeneous structure. Documents are indexed with the default \texttt{lowercase} analyzer  in a \texttt{search\_as\_you\_type} field with $4$-grams enabled.

\heading{Policy.} We use a simple DLS configuration  with a role whose index access is restricted by an intentionally simple DLS policy that enforces a public/private split:
\[
\texttt{dls}:\;\{\texttt{"term"}:\{\texttt{"public"}:\texttt{true}\}\}.
\]
This policy allows the adversary to issue queries against  the  index and observe  results and scores, but only documents with the attribute \texttt{public=true} are returned; documents with \texttt{public=false} are filtered out by DLS. The adversary can also inject documents with \texttt{public=true} into the index.

\heading{Evaluation.} Our results are shown in \Cref{tab:dls-recovery}.  
We achieve 100\% accuracy, which confirms our scoring oracles consistently distinguish
indexed and non-indexed $n$-grams. As
expected, the traversal cost of each stage scales linearly with the number of trie nodes. We reiterate that this reflects favorable conditions (e.g., single-shard index, a corpus that is static for the attack's duration) and should be read as a best case.

\begin{figure}
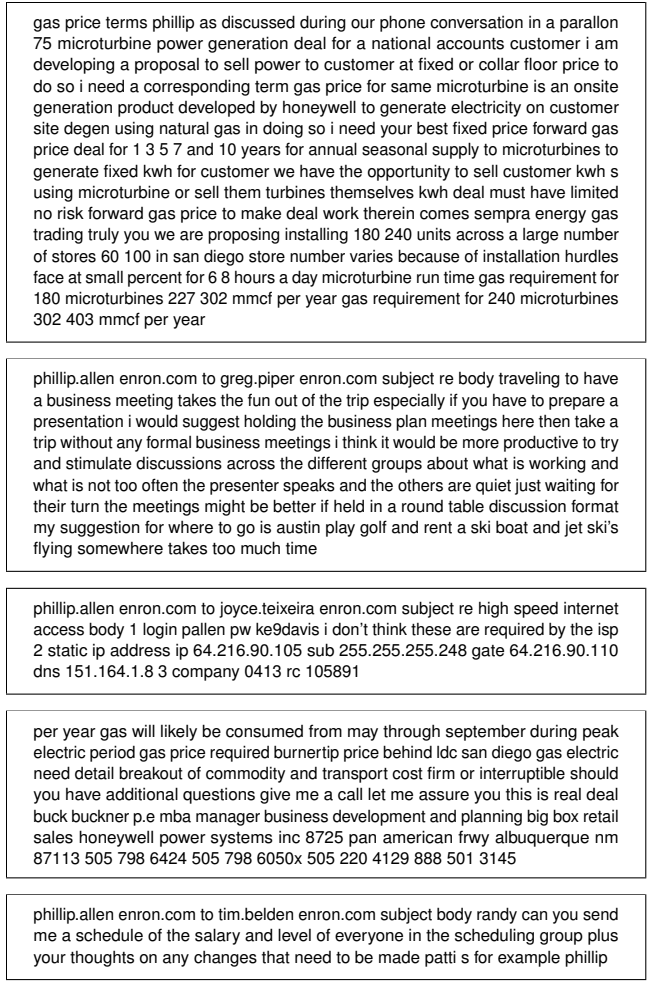

\centering \scriptsize
\begin{mdframed}
\sf gas price terms phillip as discussed during our phone conversation in a parallon 75 microturbine power generation deal for a national accounts customer i am developing a proposal to sell power to customer at fixed or collar floor price to do so i need a corresponding term gas price for same microturbine is an onsite generation product developed by honeywell to generate electricity on customer site degen using natural gas in doing so i need your best fixed price forward gas price deal for 1 3 5 7 and 10 years for annual seasonal supply to microturbines to generate fixed kwh for customer we have the opportunity to sell customer kwh s using microturbine or sell them turbines themselves kwh deal must have limited no risk forward gas price to make deal work therein comes sempra energy gas trading truly you we are proposing installing 180 240 units across a large number of stores 60 100 in san diego store number varies because of installation hurdles face at small percent for 6 8 hours a day microturbine run time gas requirement for 180 microturbines 227 302 mmcf per year gas requirement for 240 microturbines 302 403 mmcf per year
\end{mdframed}
\vspace{-10pt}
\begin{mdframed}
\sf phillip.allen enron.com to greg.piper enron.com subject re body traveling to have a business meeting takes the fun out of the trip especially if you have to prepare a presentation i would suggest holding the business plan meetings here then take a trip without any formal business meetings i think it would be more productive to try and stimulate discussions across the different groups about what is working and what is not too often the presenter speaks and the others are quiet just waiting for their turn the meetings might be better if held in a round table discussion format my suggestion for where to go is austin play golf and rent a ski boat and jet ski's flying somewhere takes too much time
\end{mdframed}
\vspace{-10pt}
\begin{mdframed}
\sf phillip.allen enron.com to joyce.teixeira enron.com subject re high speed internet access body 1 login pallen pw ke9davis i don't think these are required by the isp 2 static ip address ip 64.216.90.105 sub 255.255.255.248 gate 64.216.90.110 dns 151.164.1.8 3 company 0413 rc 105891
\end{mdframed}
\vspace{-10pt}
\begin{mdframed}
\sf per year gas will likely be consumed from may through september during peak electric period gas price required burnertip price behind ldc san diego gas electric need detail breakout of commodity and transport cost firm or interruptible should you have additional questions give me a call let me assure you this is real deal buck buckner p.e mba manager business development and planning big box retail sales honeywell power systems inc 8725 pan american frwy albuquerque nm 87113 505 798 6424 505 798 6050x 505 220 4129 888 501 3145
\end{mdframed}
\vspace{-10pt}
\begin{mdframed}
\sf phillip.allen enron.com to tim.belden enron.com subject body randy can you send me a schedule of the salary and level of everyone in the scheduling group plus your thoughts on any changes that need to be made patti s for example phillip
\end{mdframed}
\vspace{-1em}
\caption{Example $4$-gram reconstructions from $|D|=10$ Enron subset using a greedy de Bruijn  traversal~\cite{pevzner_2001}. \textbf{Takeaway:} While text is not complete, the overall meaning is preserved.}\label{figure:reconstruction-example}
\end{figure}

\heading{Document reconstruction.} We do a preliminary evaluation of  reconstructing documents from the recovered $n$-grams by building the
resulting de~Bruijn graph and then traversing the graph to recover approximate
document sequences. 
Document boundaries cannot  be uniquely
determined, but we can still extract long substrings of the original plaintext. \Cref{figure:reconstruction-example} shows  example  outputs of the de Bruijn graph traversal on the $4$-gram graph extracted from ES/OS with the default analyzer on a dataset containing $|\corp| = 10$ emails. 
While text is not complete, the meaning is preserved. Techniques from other works on $n$-gram reconstruction (e.g., \cite{galle_reconstructing_2015,hessel_how_2021}) may further improve recovery.\footnote{A natural approach is to use a large language model (LLM)
to generate more fluent reconstructions (e.g.,  
choosing among ambiguous de Bruijn paths).
We do not attempt LLM-based reconstruction  as we believe we cannot do so
fairly: the Enron corpus is a public benchmark and may appear in
the LLM's training data, which confounds attempts to determine whether actual recovery is
occurring or if the LLM has ``memorized'' the  dataset.} We leave this for future work.

\subsection{Potential Mitigations}

\vspace{-3.5pt}
\heading{Prior work.} Mitigations for BM25's DF side-channel  have been
proposed in several
works~\cite{buttcher2005security,wang_side-channel_2017,micheli2014efficient,parker2010security,singh2009search,zerr2009zerber}.
Addressing the  side-channel is relatively simple---we just need to ensure
 scores do not have a logical dependence on
private data.  The most straightforward approach is to maintain
per-user indexes, but this comes with a severe performance overhead and  reduces the effectiveness of the scoring functions. Most of the works cited above suggest approaches that
compute relevance scores using only statistics computed from user-visible
documents, but Zerr et al.~\cite{zerr2009zerber} propose adding noise to
scores. Wang et al.~\cite{wang_side-channel_2017}  
suggest either training DFs on an auxiliary public corpus or computing initial DFs
on the actual corpus, then post-filtering  non-public documents from the DFs.
Those countermeasures address the DF side-channel in \match
queries and still apply to our exact oracles. 

Mitigations for \mpp queries have not
been considered in prior work and require additional steps. Even if one uses DFs
unrelated to the corpus, the effect of expanding a term and summing the DFs will
still introduce a dependence on private documents---the contribution of a term
$t$ from the expansion set will signal that such a term exists in the private
corpus, even if the DF of $t$ is not computed on private data.

\heading{Mitigations.} A simple mitigation is to modify the computation of $\dfmpp$ to only expand
terms that are visible to a user. This can be checked quickly, once
per search.
More drastically, one could simply omit the computation of $\idfmpp$ 
from \mpp scores. When issuing a standalone \mpp query, the IDFs do not affect
the ordering of the results as the IDF is the same for every result. The IDFs only affect the scores when \emph{multiple} \mpp queries are being
combined into a single query, with scores added. Here, IDFs help promote  ``more relevant'' \mpp queries. We are not aware of this use
of \mpp, so we think many applications would not be affected by this IDF redaction.

\subsection{Discussion}

We showed that ES/OS's rich queries allow for efficient extraction
of arbitrary, unknown keywords. This  is a significant improvement over  previous work \cite{buttcher2005security,wang_side-channel_2017}, which
could only test for existence of exact terms (and not  a
prefix). Moreover,
depending on the indexing configuration, $2$, $3$, or $4$-grams are
recoverable with similar complexity, which in turn can allow for approximate text reconstruction using standard de Bruijn reconstruction techniques (depending on $\corp$'s composition).

This work does not explore the practical exploitation of this side-channel on
live systems. For example, our experiments  assume corpus document frequencies are stable for the duration of the attack, but on a live  index with concurrent writes, a term's DF (or its expansions) can change between queries and flip the oracle's decision. Over long enumerations, our adaptive strategies are more exposed to this drift. Wang et al.~\cite{wang_side-channel_2017} circumvented several
practical challenges to attack  live systems (e.g., GitHub).
In principle, their insights should transfer directly to our
setting to handle these   challenges.

\section{Related Work}
\label{sec:relatedwork}

We  already discussed prior work that studies FGAC side-channels  for RLS~\cite{dar_rls_2023,rasin_vulnerability_2024}
and DLS~\cite{buttcher2005security,wang_side-channel_2017}. Outside FGAC, there are many side-channel attacks specific to databases (e.g., the storage layer attacks from  \cite{grubbs_why_2017,espiritu_leafblower_2025,bourassa_g-dbreach_nodate,hogan_dbreach_2023,fabrega_injection_2024}). %

Other works have exploited timing side-channels in  DBMS components orthogonal to FGAC. 
Some works demonstrate timing side-channels from data-dependent execution in indexing algorithms like $\mathrm{B}^{+}$-trees~\cite{futoransky_nd2db_2007} and LSM trees~\cite{kaufman_prefix_2023}, memory deduplication  \cite{schwarzl_remote_2022}, and compression  \cite{schwarzl_practical_2023}. These techniques are slow and recover only a few bits of values per minute. Other timing attacks exploit shared microarchitectural caches \cite{shahverdi_database_2021} and filesystem synchronization~\cite{jiang_syncsync_2024,gu_i_2025} in databases.
These works differ from ours in both threat model and objective as they typically assume low-level adversarial compromise of  shared
hardware resources, whereas our attacks are purely remote and require only
query access. Finally, a set of work explores timing attacks against query-based anonymization systems \cite{boenisch_2021} and differentially private databases \cite{haeberlen_diff_2011,ratliff_2025}. Such privacy mechanisms are orthogonal to FGAC, which completely removes records from the result. 

The principle that query expressiveness leads to more leakage appears in otherwise orthogonal settings such as encrypted search (e.g., \cite{kellaris_generic_2016,falzon_full_2020,lacharite_improved_2018,grubbs_pump_2018,markatou_attacks_2023}) and statistical privacy \cite{dinur_pods_2003}. Our work adds post-filtering FGAC to this list.

\section{Conclusion}
\label{sec:conclusion}

We showed that two FGAC case studies---each with completely different technical issues---reach similar conclusions: post-filtering FGAC systems with
rich queries can fail to deliver the isolation they promise. While we propose some 
mitigations, we conclude that this architecture is inherently brittle. 
We expect that rich queries will interact badly with post-filtered FGAC
in other contexts. Of course, physically separating indexes into
per-tenant distinct structures closes the channel entirely. However, this
comes with extra overhead of managing multiple database or index instances,
which is the whole motivation for FGAC in the first place. 

Our attacks are strongest in the direct access setting, where the adversary
submits arbitrary queries to the database. While this model is realistic in some settings, as discussed in
\Cref{sec:threatmodel}, some multi-tenant systems expose only app-mediated access, where a 
trusted application exposes only a fixed set of query templates, and an
adversary can only attack via that narrow interface.
Evaluating whether reconstruction is possible under such constraints is a
valuable direction for future~work.

\vspace{-1pt}
\section*{Acknowledgments}
\vspace{-1pt}
We thank Archita Agarwal, Marilyn George, Seny Kamara,  Tarik Moataz, and Ian Ward for helpful discussions early in this work.
Generative AI tools (ChatGPT and Claude) were used to write experiment code and  to review  the text of this paper.
 This
research received no specific grant from any funding agency
in the public, commercial, or not-for-profit sectors.

\appendix
\vspace{-1pt}
\section*{Ethical Considerations}
\vspace{-1pt}
We
conducted a stakeholder-based ethics analysis  guided
by the Menlo Report \cite{menlo} and Kohno et al. \cite{kohno_ethical_2023}. For each stakeholder, we consider the benefits and harms of this research.

\heading{End users/tenants whose data is protected by FGAC.} The main risk is unauthorized inference about protected rows or documents if their service deploys a vulnerable post-filtering design. The benefit is that operators  receive clearer guidance for reducing this risk. We mitigated risks by experimenting in controlled testbeds and
not on third-party deployments. 

\heading{Developers and organizations using FGAC systems.} The risk is privacy, compliance, and financial harm if confidentiality boundaries fail, as well as operational cost to reassess query interfaces, RLS/DLS configurations, indexing strategies, logging, and monitoring. The benefit is improved ability to assess whether their deployments expose the relevant query features that make them vulnerable to the attack. 
To help operators assess applicability, we explicitly state attack prerequisites  (e.g., required query features, score visibility, injection capability, indexing choices), though we acknowledge that some mitigations  may impose significant
 tradeoffs. 

\heading{Database  maintainers.} The risk is triage burden and possible reputation impact. The benefit is a more precise understanding of the leakage of expressive query interfaces. 
To  mitigate risks, we disclosed our work on January 25, 2026 to  PostgreSQL, Elastic, and OpenSearch, and proposed a 90-day window. We had received acknowledgments from all  parties by February 9.
Elastic indicated they consider the vulnerabilities as low priority at this time. OpenSearch and PostgreSQL sent a basic acknowledgment. At camera-ready time, we were not aware of any patch release dependency that conflicted with the  publication timeline, so we proceeded with publication. 

\heading{The research community and authors.} The security research
community benefits from a clearer understanding of the security of FGAC mechanisms. At the same time, attackers
 benefit from publication by learning how to implement more effective attacks. We reduced this risk by notifying affected parties before publication, emphasizing mitigations and limitations, and avoiding live  testing.
The research authors are also stakeholders, but we do not believe that a research team can   unbiasedly assess  the harms and benefits of their own research on themselves, so we abstain from this analysis.

\heading{Decision to research and to publish.}
Our approach to this work was based on a structured comparison between the  risks  and  benefits of public disclosure. We considered (1) not publishing, (2) private, vendor-only disclosure, or (3) disclosing  and publishing publicly.
We  concluded that vendor-only disclosure would be insufficient, as we believe effective mitigation requires case-by-case configuration changes, restrictions on query interfaces, and  risk assessment---choices  made by application developers  rather than by database vendors.

We then considered the risk that publication could help adversaries. We mitigated this risk in several ways: we evaluated only systems under our administrative control, did not test against production services or private user data, and disclosed  to the evaluated affected parties before public release.

Finally, we considered broader ethical standards. From a \textit{Beneficence} perspective, 
publication provides better understanding
of when FGAC threat models fail under expressive query interfaces. This enables more secure
engineering guidance and makes the limits of FGAC clearer.  From a \textit{Justice} perspective,  publication shifts the burden of discovering subtle side-channels away from individual customers and users, whose data might be protected by such systems.
From  a \textit{Respect for Persons} perspective, it advances users' privacy interests by motivating deployment of  mitigations. From a \textit{Respect for Law and Public Interest} perspective, we avoided attacking live systems and engaged in coordinated disclosure. 

On balance, we concluded that  disclosure \textit{and} publication best serve the public interest. The  practical severity of composing side-channel mechanisms   with expressive query interfaces was not previously understood, and the  design pattern is not  implementation-specific and thus may recur in other systems. This chain of reasoning led us to publish this paper.

\vspace{-1pt}
\section*{Open Science}
\vspace{-1pt}
Our attack and experiment setup scripts are available at \url{https://doi.org/10.5281/zenodo.20370280}.
\vspace{-1pt}

\bibliography{rls}
\bibliographystyle{abbrvurl}

\section{RLS Prefix Queries}
\label{appendix:prefix-queries}

For string attributes, we implement prefix tests using \emph{range predicates} rather than
PostgreSQL's pattern-matching operators.
Although PostgreSQL provides dedicated prefix operators (e.g., \texttt{LIKE} and
\texttt{ILIKE}), these are not marked \texttt{LEAKPROOF} in general (e.g., they can raise
errors under certain collations/encodings), and therefore cannot be reliably pushed down past
the RLS security barrier. In contrast, standard comparison operators on text
(\texttt{>=}, \texttt{<}) are \texttt{LEAKPROOF} and can be used to express the same prefix test
in a way that the optimizer can apply using a B$^+$-tree index on the string attribute.

\heading{Collation and ordering.}
In PostgreSQL, comparisons on \texttt{text}/\texttt{varchar} are performed under the
column's \emph{collation}~\cite{noauthor_232_2025}, which defines the total order used by B$^+$-tree index scans.
Our range-based prefix formulation assumes a fixed, known collation for the target column,
and that the adversary can compute the successor bound $p_{\mathsf{next}}$ with respect to
that same order (e.g., by treating values as byte strings under the \texttt{C} collation).
If a locale-dependent collation is used, computing $p_{\mathsf{next}}$ may require access to
the same collation rules; in such deployments, the attacker can instead target columns with
binary/bytewise order (or normalize text into a deterministic byte representation).

\heading{Attack formulation.}
Fix a string attribute \texttt{attr} and a prefix $p$.
Let $p_{\mathsf{next}}$ denote the smallest string strictly greater than $p$ such that every
string beginning with prefix $p$ lies in the half-open interval $[p, p_{\mathsf{next}})$ under
lexicographic order.\footnote{$p_{\mathsf{next}}$ can be computed by incrementing
the last character of $p$ within a chosen finite alphabet (or by using a standard ``next
lexicographic string'' routine) and treating strings as byte sequences under a fixed collation.
Our attack assumes the adversary knows the ordering used by the database index.}
Then the predicate ``\texttt{attr} begins with $p$'' is equivalent to:
\[
p \;\le\; \texttt{attr} \;<\; p_{\mathsf{next}}.
\]
Accordingly, we issue oracle probes of the form:
\begin{lstlisting}[language=SQL]
SELECT 1 FROM patients WHERE attr >= p AND attr < p_next LIMIT 1;
\end{lstlisting}

Then, to enumerate unknown strings, the adversary starts from the empty prefix and extends it
character-by-character.
Given a current prefix $p$, for each character $c$ in an assumed alphabet $\Sigma$ the
adversary tests the extended prefix $p\Vert c$ using the range probe above.
Any extension that yields a positive oracle response is retained and further extended.
The process terminates for a prefix when no extension is positive, indicating that no longer
string in the table begins with that prefix.
In practice, the number of probes depends on the number of distinct strings present (and their
shared prefixes), rather than on the size of the overall string universe.

\end{document}